\documentclass[showpacs,aps,prd,reprint,superscriptaddress,nofootinbib,longbibliography,preprintnumbers]{revtex4-1}
\usepackage[colorlinks=true, pdfstartview=FitV, linkcolor=magenta,citecolor=blue, urlcolor=magenta,
bookmarks=true, bookmarksnumbered=true, breaklinks]{hyperref}
\usepackage[dvipdfmx]{graphicx}
\usepackage{here}
\usepackage{url}
\usepackage{amsmath,amssymb,bm,color,longtable,mathrsfs,amsfonts,slashed}

\def\blue{\color{blue}}
\newcommand{\Slash}[1]{{\ooalign{\hfil/\hfil\crcr$#1$}}}

\usepackage{ulem}

\begin{document}
\begin{flushright}
\end{flushright}


\title{Axial anomaly effect on three-quark and five-quark singly heavy baryons}

\author{Hiroto~Takada}
\email[]{takada@hken.phys.nagoya-u.ac.jp}
\affiliation{Department of Physics, Nagoya University, Nagoya 464-8602, Japan}

\author{Daiki~Suenaga}
\email[]{daiki.suenaga@riken.jp}
\affiliation{Nishina Center for Accelerator-Based Science, RIKEN, Wako 351-0198, Japan}
\affiliation{Research Center for Nuclear Physics (RCNP), Ibaraki, Osaka 567-0047, Japan}

\author{Masayasu~Harada}
\email[]{harada@hken.phys.nagoya-u.ac.jp}
\affiliation{Department of Physics, Nagoya University, Nagoya 464-8602, Japan}
\affiliation{Kobayashi-Maskawa Institute for the Origin of Particles and the Universe, Nagoya University, Nagoya, 464-8602, Japan}
\affiliation{Advanced Science Research Center, Japan Atomic Energy Agency (JAEA), Tokai 319-1195, Japan}

\author{Atsushi~Hosaka}
\email[]{hosaka@rcnp.osaka-u.ac.jp}
\affiliation{Research Center for Nuclear Physics (RCNP), Ibaraki, Osaka 567-0047, Japan}
\affiliation{Advanced Science Research Center, Japan Atomic Energy Agency (JAEA), Tokai 319-1195, Japan}

\author{Makoto~Oka}
\email[]{makoto.oka@riken.jp}
\affiliation{Nishina Center for Accelerator-Based Science, RIKEN, Wako 351-0198, Japan}
\affiliation{Advanced Science Research Center, Japan Atomic Energy Agency (JAEA), Tokai 319-1195, Japan}

\date{\today}

\begin{abstract}
Effects of the $U(1)_A$ axial anomaly on the mass spectrum of singly heavy baryons (SHBs) is studied in terms of the chiral effective theory based on the chiral linear representation for light flavors. We consider SHBs made of both three quarks ($Qqq$) and five quarks ($Qqqq\bar{q}$). For the three-quark SHBs we prove that the inverse mass hierarchy for the negative-parity $\Lambda_c$ and $\Xi_c$ is realized only when the $U(1)_A$ anomaly is present. For the five-quark SHBs, in contrast, it is found that the $U(1)_A$ anomaly does not change the mass spectrum at the leading order, and accordingly their decay properties induced by emitting a pseudoscalar meson are not affected by the anomaly. Moreover, taking into account small mixings between the three-quark and five-quark SHBs, we find that the observed $\Xi_c$ excited state, either $\Xi_c(2923)$ or $\Xi_c(2930)$, can be consistently regarded as a negative-parity SHB that is dominated by the five-quark component. We also predict a new negative-parity five-quark dominant $\Lambda_c$, whose mass is around $2700$ MeV and the decay width is of order a few MeV, which provides useful information for future experiments to check our description.
 \end{abstract}

\pacs{}

\maketitle

\section{Introduction}
\label{sec:Introduction}

Chiral symmetry for light-flavor ($u$, $d$ and $s$) quarks is one of the important symmetries of quantum chromodynamics (QCD). In fact, the spontaneous breakdown of chiral symmetry enables us to understand the mass generation of hadrons from almost massless light quarks~\cite{cheng1984gauge}, and simultaneously enables us to describe the low-energy dynamics for the associated Nambu-Goldstone (NG) bosons (such as pions) systematically~\cite{Gasser:1983yg,Gasser:1984gg}. Another significant symmetry property of QCD is the $U(1)_A$ axial anomaly~\cite{Bell:1969ts,Adler:1969gk}, i.e., nonconservation of the $U(1)_A$ axial charges induced by instantons~\cite{tHooft:1986ooh}, which is essential to explain a large mass of $\eta'$ meson. 

In view of the above symmetry aspects, the studies based on chiral effective-model approaches have been broadly carried out for light mesons and baryons. In addition to hadrons including only light flavors, heavy-light mesons composed of one heavy quark ($c$ or $b$ quark) and one light quark as well as doubly heavy baryons of two heavy quarks and one light quark have also been explored within the chiral models~\cite{Nowak:1992um,Bardeen:1993ae,Bardeen:2003kt,Ma:2015lba,Ma:2015cfa,Ma:2017nik}.

Since the heavy quark plays a role of a spectator due to its large mass, studies of the heavy hadrons allow us to extract information on the QCD symmetry properties carried by light quarks despite being confined~\cite{Neubert:1993mb,manohar2000heavy}. In other words, such open heavy hadrons provide us with useful testing ground toward understanding dynamics of light-quark clusters which are not color-singlet. From those examinations for various flavor system, it is expected that our insights into the mechanism of flavor dependent or independent hadron mass generations would be deepened.

In this regard, singly heavy baryons (SHBs), which are composed of one heavy quark and one light {\it diquark}, serve as another useful probe to unveil the dynamics of color-nonsinglet objects~\cite{Hong:2004ux}. That is, the diquark dynamics stemming from chiral symmetry and the $U(1)_A$ axial anomaly is reflected to the mass and decay properties of SHBs. Theoretical studies of SHBs focusing on the diquarks have been done from chiral  models~\cite{Ebert:1995fp,Kawakami:2018olq,Kawakami:2019hpp,Harada:2019udr,Dmitrasinovic:2020wye,Kawakami:2020sxd}, quark models~\cite{Copley:1979wj,Yoshida:2015tia}, and diquark-heavy-quark potential descriptions~\cite{Kim:2020imk,Kim:2021ywp,Kim:2022pyq}. Accordingly, the spectroscopy of SHBs are being energetically explored experimentally at, e.g., SLAC, KEK, and LHC. In addition, the chiral-partner structures of the SHBs at high temperature based on a chiral model of diquarks has also been examined in Ref.~\cite{Suenaga:2023tcy}.

In Ref.~\cite{Suenaga:2021qri}, a 5-quark picture ($Qqq\bar{q}q$) was proposed to describe the so-called Roper-like 
baryons, $\Lambda_c(2765)$ and $\Xi_c(2970)$.
Using the linear representation of chiral symmetry, the sequential decays of the 5-quark SHBs induced by emitting two NG bosons were reasonably explained~\cite{Suenaga:2022ajn}. The chiral representation of the 5-quark SHBs is identical to that of the 3-quark ones but their axial charges are different. Hence, classification of them from the $U(1)_A$ axial charges is inevitable to understand the distinction of symmetry properties between the two types of SHBs. Moreover, in Ref.~\cite{Harada:2019udr} it was found that the $U(1)_A$ anomaly effects can lead to the so-called {\it inverse mass hierarchy} where $\Lambda_c$ becomes heavier than $\Xi_c$ for negative-parity 3-quark SHBs. This implies that the anomaly plays significant roles in the mass spectrum of SHBs.

Motivated by the above observations, in this paper, we examine influences of the $U(1)_A$ axial anomaly on the mass spectrum and decay properties of the SHBs based on 3-quark and 5-quark pictures. After such considerations, we show our predictions of the masses and decay widths of the negative-parity 5-quark dominant $\Lambda_c$ baryon.

This paper is organized as follows. In Sec.~\ref{sec:model}, we present our effective Lagrangian including the 3-quark and 5-quark SHBs based on $SU(3)_L\times SU(3)_R$ chiral symmetry, and explanations of contributions from the $U(1)_A$ axial anomaly are provided with referring to quark-line diagrams. In Sec.~\ref{sec:3-quark}, influences of the anomaly on mass spectrum and decay widths of the pure 3-quark SHBs are investigated in detail, and similar considerations for the 5-quark SHBs are provided in Sec.~\ref{sec:5-quark}. In Sec.~\ref{sec:35Mixing}, mixings between the 3-quark and 5-quark SHBs are incorporated and we present predictions of the negative-parity 5-quark dominant $\Lambda_c$ baryon. In Sec.~\ref{sec:DecayLambdaC}, we provide discussions on the predicted $\Lambda_c$ baryon. Finally in Sec.~\ref{sec:Conclusions}, we conclude the present study.

\section{Model}
\label{sec:model}

In this section, we present our effective model for the SHBs based on chiral symmetry of the diquarks.

In order to describe both the ground state SHBs, $\Lambda_c(2286)$ and $\Xi_c(2470)$, and low lying excited states such as the Roper like ones, $\Lambda_c(2765)$ and $\Xi(2970)$, from chiral symmetry point of view, we introduce four diquarks $d_R$, $d_L$, $d'_R$ and $d_L'$ whose quark contents are given by~\cite{Suenaga:2021qri}
\begin{eqnarray}
     {(d_R)}^\alpha_a&\sim&\epsilon_{abc}\epsilon^{\alpha\beta\gamma} (q^{T}_{R})_b^\beta C(q_R)_c^\gamma \  , \nonumber\\
     {(d_L)}^\alpha_i&\sim&\epsilon_{ijk}\epsilon^{\alpha\beta\gamma}(q^{T}_{L})_j^\beta C(q_L)_k^\gamma\ , \nonumber\\
   {(d^{\prime}_R)}^\alpha_i&\sim&\epsilon_{abc}\epsilon^{\alpha\beta\gamma}{(q^T_R)}^\beta_b C{(q_R)}^\gamma_c[{(\bar{q}_L)}^\delta_i{(q_R)}^\delta_a] \ , \nonumber\\
    {(d^{\prime}_L)}^\alpha_a&\sim&\epsilon_{ijk}\epsilon^{\alpha\beta\gamma}{(q^T_L)}^\beta_j C{(q_L)}^\gamma_k[{(\bar{q}_R)}^\delta_a{(q_L)}^\delta_i]\ . \label{Diquarks}
\end{eqnarray}
In this equation, $q_{R(L)} = \frac{1\pm\gamma_5}{2}q$ is the right-handed (left-handed) quark field. The subscripts ``$a,b, \cdots$'' and ``$i, j, \cdots$'' denote right-handed and left-handed chiral indices, respectively, and the superscripts ``$\alpha, \beta, \cdots$'' stands for color indices. The $4\times4$ matrix $C=i\gamma^2\gamma^0$ is the charge-conjugation Dirac matrix. Thus, while $d_R$ and $d_L$ are the conventional diquarks consisting of two quarks, $d_R'$ and $d_L'$ are regarded as the {\it tetra-diquarks} made of three quarks and one antiquark. The chiral representation of $d_R$, $d_L$, $d^{\prime}_R$, $d^{\prime}_L$ reads
\begin{eqnarray}
    d_R\sim{(\bm 1,\bar{\bm 3})}_{+2} \ , \ \  d_L\sim{(\bar{\bm 3},\bm 1)}_{-2}\ ,  \nonumber\\
    d^{\prime}_R\sim {(\bar{\bm 3},\bm 1)}_{+4} \ , \ \    d^{\prime}_L\sim {(\bm 1,\bar{\bm 3})}_{-4} \ , \label{ChiralRep}
\end{eqnarray}
where the subscripts, e.g., $+2$ for $d_R$, represent the $U(1)_A$ axial charge carried by the diquarks. Equation~(\ref{ChiralRep}) shows that the axial charges of the tetra-diquarks are distinct from those of the conventional ones, which allows us to distinguish the two types of diquarks, although $d_R$ and $d_L'$ ($d_L$ and $d_R'$ ) belong to the identical chiral representation.

The interpolating fields of SHBs are given by attaching a heavy quark $Q$ to the diquark as
\begin{eqnarray}
    B_{R,a}\sim Q^\alpha{(d_R)}^\alpha_a\ ,\ \  B_{L,i}\sim Q^\alpha{(d_L)}^\alpha_i\ , \nonumber\\
    B^{\prime}_{R,i}\sim Q^\alpha{(d^{\prime}_R)}^\alpha_i\ , \ \  B^{\prime}_{L,a}\sim Q^\alpha{(d^{\prime}_L)}^\alpha_a\ . \label{SHBs}
\end{eqnarray}
From this definition, one can see that $B_{R (L)}$ and $B'_{R(L)}$ are regarded as a 3-quark state and 5-quark state, respectively. Besides, Eq.~(\ref{ChiralRep}) implies that chiral transformation laws of the SHBs read
\begin{eqnarray}
B_{R}\rightarrow B_{R}g_R^{\dagger}\ , \ \  B_{L}\rightarrow B_{L}g_L^{\dagger}\ ,\nonumber\\
B^{\prime}_{R}\rightarrow B^{\prime}_{R}g_L^{\dagger}\ , \ \  B^{\prime}_{L}\rightarrow B^{\prime}_{L}g_R^{\dagger}\ , \label{BChiral}
\end{eqnarray}
with $g_{R(L)}\in SU(3)_{R(L)}$. It should be noted that the SHBs in Eq.~(\ref{SHBs}) are heavy-quark spin-singlet (HQS-singlet) of spin $1/2$ since the diquarks in Eq.~(\ref{Diquarks}) are Lorentz scalar.

From Eq.~(\ref{BChiral}), an effective Lagrangian describing the 3-quark SHBs and 5-quark SHBs coupling with light mesons which is invariant under $SU(3)_L\times SU(3)_R$ transformation is constructed as
\begin{eqnarray}
{\cal L}_{\rm SHB} = \mathcal{L} _{3q}+\mathcal{L}_{5q}+\mathcal{L} _{\rm mix} \ , \label{LSHB}
\end{eqnarray}
where
\begin{eqnarray}
{\cal L}_{3q} &=& \sum_{\chi =L,R}(\bar{B}_\chi iv\cdot\partial B_\chi - \mu _1 \bar{B}_\chi B_\chi) \nonumber\\
&-&\frac{\mu_3}{f_\pi^{2}}\Big[\bar{B}_{L}(\Sigma\Sigma ^{\dagger})^TB_{L} + \bar{B}_{R}(\Sigma ^{\dagger}\Sigma)^TB_{R}\Big] \nonumber\\
    &-&\frac{g_1}{2f_\pi} \left(\epsilon_{ijk}\epsilon_{abc}  \bar{B}_{L,k}\Sigma_{ia}\Sigma_{jb}B_{R,c}+ {\rm h.c.} \right) \nonumber\\
    &-&g_1^{\prime}(\bar{B}_L\Sigma^{\ast}B_R+{\rm h.c.})\ , \label{L3q}
\end{eqnarray}
\begin{eqnarray}
{\cal L}_{5q} &=& \sum_{\chi =L,R}(\bar{B}'_\chi iv\cdot\partial B'_\chi - \mu _2 \bar{B}'_\chi B'_\chi) \nonumber\\ 
&-&\frac{\mu_4}{f_\pi^{2}}\Big[ \bar{B}'_{R} (\Sigma\Sigma^{\dagger})^TB'_{R}+\bar{B}'_L(\Sigma^{\dagger}\Sigma)^TB'_{L} \Big] \nonumber\\
&-&\frac{g_2}{6f_\pi^3}\Big[(\epsilon_{abc}\epsilon_{ijk}\Sigma^{\dagger}_{ci}\Sigma^{\dagger}_{bj}\Sigma^{\dagger}_{ak})(\bar{B}^{\prime}_{R}\Sigma^*B^{\prime}_{L})+{\rm h.c.}\Big] \nonumber\\
&-&\frac{g_3}{2f_\pi^3}\left(\epsilon_{abc}\epsilon_{ijk}\bar{B}^{\prime}_{R,l}{\Sigma^{\dagger}_{c l}}{\Sigma^{\dagger}_{b i}}{\Sigma^{\dagger}_{a j}}{\Sigma^{\dagger}_{d k}}B^{\prime}_{L,d}+{\rm h.c.} \right) \nonumber\\
    &+&g_2^{\prime} \left(\bar{B}^{\prime}_R\Sigma^{\ast}B^{\prime}_L+\bar{B}^{\prime}_L\Sigma^{T}B^{\prime}_R\right) \ ,  \label{L5q}
\end{eqnarray}
and
\begin{eqnarray}   
{\cal L}_{\rm mix} &=& -\mu_1^{\prime}(\bar{B}_R B'_L +\bar{B}'_L B_R +\bar{B}_L B'_R +\bar{B}'_R B_L) \nonumber\\
&-&g_4(\bar{B}^{\prime}_R\Sigma^{\ast}B_R+\bar{B}_L\Sigma^{\ast}B^{\prime}_L + \mathrm{h.c.}) \ .      \label{LMix}
\end{eqnarray}
In these equations, $\Sigma$ is a light meson nonet which belongs to
\begin{eqnarray} 
\Sigma \sim ({\bm 3},\bar{\bm 3})_{-2}\ ,
\end{eqnarray}
or more explicitly, $\Sigma$ transforms under the $SU(3)_L\times SU(3)_R$ chiral transformation as 
\begin{eqnarray}
\Sigma \to g_L\Sigma g_R^{\dagger}\ .
\end{eqnarray}
The dimensionless quantity $v$ in the Lagrangian stands for the velocity of the SHB\sout{s}. In Eqs.~(\ref{L3q}) -~(\ref{LMix}), chiral symmetry properties of the contributions including the antisymmetric tensor are rather obscure, so here we provide an explanation of their chiral invariance, by focusing on the $g_1$ term in Eq.~(\ref{L3q}) as an example. As for this term, all of the subscripts $i$, $j$ and $k$ in $\Sigma_{ia}$, $\Sigma_{jb}$ and $\bar{B}_{L,k}$ denote indices of the ${\bm 3}$ representation of left-handed $SU(3)_L$ group, and hence, by contracting these indices with the antisymmetric tensor $\epsilon_{ijk}$, one obtains an $SU(3)_L$ chiral-singlet piece. Likewise, the indices $a$, $b$ and $c$ in $\Sigma_{ia}$, $\Sigma_{jb}$ and ${B}_{R,c}$ belong to the $\bar{\bm 3}$ representation of $SU(3)_R$, so the contraction with $\epsilon_{abc}$ leaves an $SU(3)_R$ chiral-singlet. As a result, chiral invariance of the term becomes manifest. Our Lagrangian possesses $SU(2)_h$ heavy-quark spin symmetry (HQSS) as well as $SU(3)_L\times SU(3)_R$ chiral symmetry, which can be easily understood by a fact that it does not include any Dirac $\gamma^\mu$ matrices~\cite{Neubert:1993mb,manohar2000heavy}.

\begin{figure}[t]
\centering
\includegraphics*[scale=0.59]{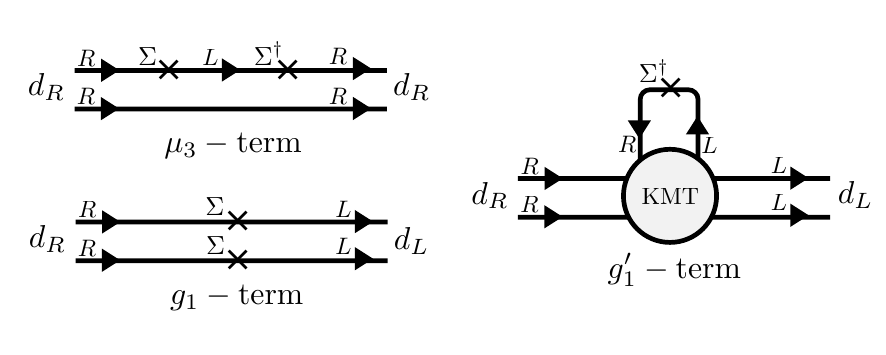}
  \caption{Quark line diagrams for each term of the Lagrangian~(\ref{L3q}). The heavy quark is a spectator and omitted here.}
\label{fig:L3}
\end{figure}

\begin{figure}[t]
\centering
\includegraphics*[scale=0.56]{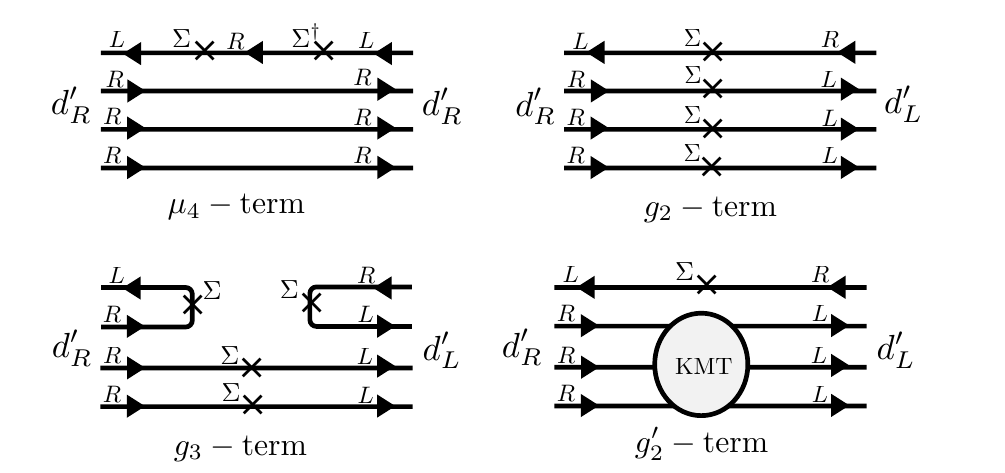}
  \caption{Quark line diagrams for each term of the Lagrangian~(\ref{L5q}). }
\label{fig:L5}
\end{figure}

\begin{figure}[t]
\centering
\includegraphics*[scale=0.59]{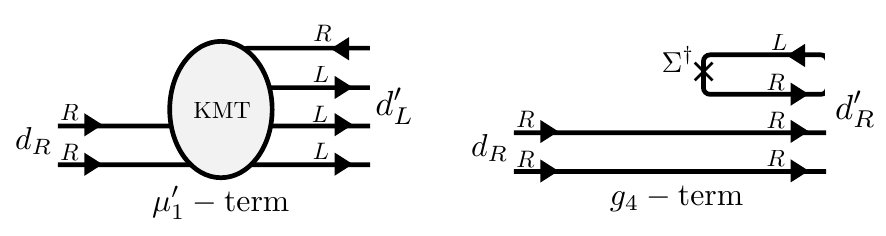}
  \caption{Quark line diagrams for each term of the Lagrangian~(\ref{LMix}).}
\label{fig:LMix}
\end{figure}

Our counting scheme in constructing the Lagrangian~(\ref{LSHB}) is as follows: First, we have written down all possible terms invariant under the $U(1)_A$ axial transformation in addition to the $SU(3)_L\times SU(3)_R$ chiral transformation with the smallest number of $\Sigma^{(\dagger)}$. Next, we have included leading terms which break only the $U(1)_A$ axial symmetry. Because of these reasonings, the $g_2$ and $g_3$ terms in Eq.~(\ref{L5q}) containing four $\Sigma^{(\dagger)}$'s, which at first glance seems to be higher order, are present. In fact, the $\mu_1$, $\mu_2$, $\mu_3$, $\mu_4$, $g_1$, $g_2$, $g_3$ and $g_4$ terms are invariant under the $U(1)_A$ axial transformation, whereas the remaining $g'_1$, $g'_2$ and $\mu_1'$ terms violate the $U(1)_A$ axial symmetry. That is, only the latter three contributions are responsible for the $U(1)_A$ axial anomaly. It should be noted that a trace of $\Sigma^\dagger\Sigma$ which is not directly connected to quark lines inside the SHBs, e.g., ${\rm tr}[\Sigma^\dagger\Sigma](\bar{B}_L B_{L}+\bar{B}_RB_R)$ term, can be also included within our present counting rule. But, such contributions are ignored in our present analysis since they do not essentially affect mass spectrum and one-pseudoscalar-meson emission decays of the SHBs.

In order to gain insights into the $U(1)_A$ axial properties of the contributions, we depict quark-line diagrams of each interaction term in Figs.~\ref{fig:L3} -~\ref{fig:LMix}. Figure~\ref{fig:L3} shows that chirality flips induced by $\Sigma^{(\dagger)}$ occur twice in one quark line for $\mu_3$ term, and such flips occur in two quark lines for $g_1$ term. As a result, $U(1)_A$ symmetry for these two terms becomes manifest since all right-handed and left-handed quark lines are preserved. Meanwhile, as displayed in the figure, $g_1'$ term includes the so-called Kobayashi-Maskawa-'t Hooft (KMT) six-point interaction~\cite{Kobayashi:1970ji,Kobayashi:1971qz,tHooft:1976snw,tHooft:1976rip} which leads to the chirality nonconservation representing the $U(1)_A$ axial anomaly.

As for the tetra-diquarks, from Fig.~\ref{fig:L5} one can see that the double chirality flip occurs for the antiquark line in the $\mu_4$ term since the chiral indices of the diquark are carried by the antiquark as in Eq.~(\ref{Diquarks}), which is distinct from the $\mu_3$ term despite the identical coupling structure at the Lagrangian level. Besides, in the $g_2$ term, not only the antiquark line but also the remaining three quark lines interact with $\Sigma^{(\dagger)}$ to flip their chiralities, in which the antisymmetric-tensor structures of the latter three quarks are directly connected to those of $\epsilon_{abc}\epsilon_{ijk}\Sigma^{\dagger}_{ci}\Sigma^{\dagger}_{bj}\Sigma^{\dagger}_{ak}$ piece. Meanwhile, the $g_3$ term includes contributions where one antiquark line is connected to another quark line through $\Sigma^{(\dagger)}$, since, for instance, the chiral index of $B'_{L,d}$ is related to $\Sigma_{dk}^\dagger$ having a contraction with other meson fields by $\epsilon_{ijk}$. The quark line for the last $g_2'$ term is simply understood by replacing the $\epsilon_{abc}\epsilon_{ijk}\Sigma^{\dagger}_{ci}\Sigma^{\dagger}_{bj}\Sigma^{\dagger}_{ak}$ piece in the $g_2$ term by the KMT interaction, which manifestly shows the chirality nonconservation and the $U(1)_A$ axial anomaly effects.

The diagrams for the mixing terms depicted in Fig.~\ref{fig:LMix} are rather simple. In the $\mu_1'$ term, the mixing between the conventional diquark and tetra-diquark is supplemented by the anomalous KMT interaction, and in the $g_4$ term such a mixing is  simply provided by $\Sigma^{(\dagger)}$ within the tetra-diquark.

Under the spontaneous breaking of chiral symmetry, $\Sigma$ acquires vacuum expectation values (VEVs) of the form 
\begin{eqnarray}
\langle\Sigma\rangle=f_{\pi}\mathrm{diag}(1,1,A)\ , \label{VEV}
\end{eqnarray}
where the parameter $A$ incorporates a violation of $SU(3)_{L+R}$ flavor symmetry due to the presence of a large $s$ quark mass. In our present analysis, we take $f_{\pi}=$93 MeV and $A=\frac{2f_K-f_{\pi}}{f_{\pi}}=1.38$ (hence $f_K=111$ MeV). Replacing $\Sigma^{(\dagger)}$ by its VEVs~(\ref{VEV}) in our model~(\ref{LSHB}), masses of the SHBs are evaluated.

In the following sections, we present our results of the analyses of our effective Lagrangian. We first switch off the mixing term, Eq.~(\ref{LMix}), in Secs.~\ref{sec:3-quark} and~\ref{sec:5-quark}, so as to explore influences of the $U(1)_A$ axial anomaly on the mass spectrum of the 3-quark SHBs and the 5-quark SHBs separately.
Then, in Sec.~\ref{sec:35Mixing}, we revive the mixing to investigate the full spectrum and decay properties of SHBs.

\section{Analysis of 3-quark SHBs}
\label{sec:3-quark}

Here, we investigate the masses and decay widths of SHBs which contain only 3-quark states from Eq.~(\ref{L3q}) in the absence of mixing effects~(\ref{LMix}). 

Flavor basis of the SHBs is obtained by the diagonal components of $SU(3)_L$ and $SU(3)_R$ groups, i.e., by putting $i=a$ in the interpolating fields~(\ref{SHBs}). Then, from Eq.~(\ref{SHBs}) together with Eq.~(\ref{Diquarks}), one can find that parity eigenstates of the 3-quark SHBs are obtained as linear combinations of $B_R$ and $B_L$ as
\begin{eqnarray}
 B_{\pm,i}=\frac{1}{\sqrt{2}}(B_{R,i}\mp B_{L,i})\ , \label{B3Eigen}
\end{eqnarray}
where the sign of $B_{\pm,i}$ in the left-hand side (LHS) represents the parity. Accordingly, mass eigenvalues of the 3-quark SHBs read
\begin{eqnarray}
M[{\Lambda}_c^{[\bm3]}(\pm)] &=& m_B + \mu_1+\mu_3\mp f_{\pi}(g_1+Ag_1')\ , \nonumber\\
M[{\Xi}_c^{[\bm3]}(\pm)] &=& m_B +  \mu_1+A^2\mu_3\mp f_{\pi}(Ag_1+g_1') \ .  \nonumber\\
\label{Mass3Q}
\end{eqnarray}
In this equation, ${\Xi}_c^{[\bm3]}(\pm)$ and ${\Lambda}_c^{[\bm3]}(\pm)$ are the SHBs composed of $suc$ ($sdc$) and $udc$ carrying the parity $\pm$, respectively, where the superscript $[{\bm 3}]$ is shown to emphasize that they are 3-quark SHBs. The quantity $m_B$ is a mass parameter introduced to defined a heavy-baryon effective theory~\cite{Neubert:1993mb,manohar2000heavy}, so that we can choose its value arbitrarily. Equation~(\ref{Mass3Q}) indicates that, when we focus on $M[{\Lambda}_c^{[\bm3]}(\pm)]$, $\langle\bar{s}s\rangle$ contributions denoted by $A$ is incorporated into the mass through the anomalous $g_1'$ term although $\Lambda_c^{[{\bm 3}]}$ does not contain the $s$-quark content. Such peculiar structure is understood by the KMT interaction as displayed in Fig.~\ref{fig:L3} which mixes all flavors $u$ ($\bar{u}$), $d$ ($\bar{d}$) and $s$ ($\bar{s}$).

The positive-parity SHBs $\Lambda_c^{[\bm3]}(+)$ and $\Xi_c^{[\bm3]}(+)$ correspond to the experimentally observed ground-state $\Lambda_c(2286)$ and $\Xi_c(2470)$, and hence we use their masses as inputs~\cite{Workman:2022ynf}:
\begin{eqnarray}
&& M[{\Lambda}_c^{[\bm3]}(+)] = 2286\, {\rm MeV}\ , \nonumber\\
&& M[{\Xi}_c^{[\bm3]}(+)] = 2470\, {\rm MeV} \ , \label{GSInput}
\end{eqnarray}
which allows us to fix two of the parameters $\mu_1$, $\mu_3$, $g_1$ and $g_1'$ in Eq.~(\ref{Mass3Q}).\footnote{As explained below Eq.~(\ref{Mass3Q}), $m_B$ is not a model parameter to be fixed, but we can determine freely. In fact, the $m_B$ dependence can be absorbed into $\mu_1$.} As for the unobserved negative-parity SHBs, we assume that their masses are larger than the positive-parity ones sharing the same flavor contents: 
\begin{eqnarray}
&& M[{\Lambda}_c^{[\bm3]}(-)] > M[{\Lambda}_c^{[\bm3]}(+)]\ , \nonumber\\
&& M[{\Xi}_c^{[\bm3]}(-)]> M[{\Xi}_c^{[\bm3]}(+)]\ , \label{3QBound}
\end{eqnarray}
since the negative-parity SHBs are regarded as orbitally excited states.\footnote{Note that the experimentally observed states, $\Lambda_c(2595) (J^P=1/2^-)$ and its flavor partner~\cite{Workman:2022ynf}, are not chiral-partner states that we concern here. In a quark-model description, $\Lambda_c(2595)$ is regarded as the so-called $\lambda$-mode excited baryon since being the ground state of $J^P=1/2^-$. Thus, the chiral-partner state which corresponds to the $\rho$-mode excited baryon must be heavier than $\Lambda_c(2595)$~\cite{Yoshida:2015tia}.}

\begin{figure}[t]
\centering
\hspace*{-0.4cm}
\includegraphics*[scale=0.48]{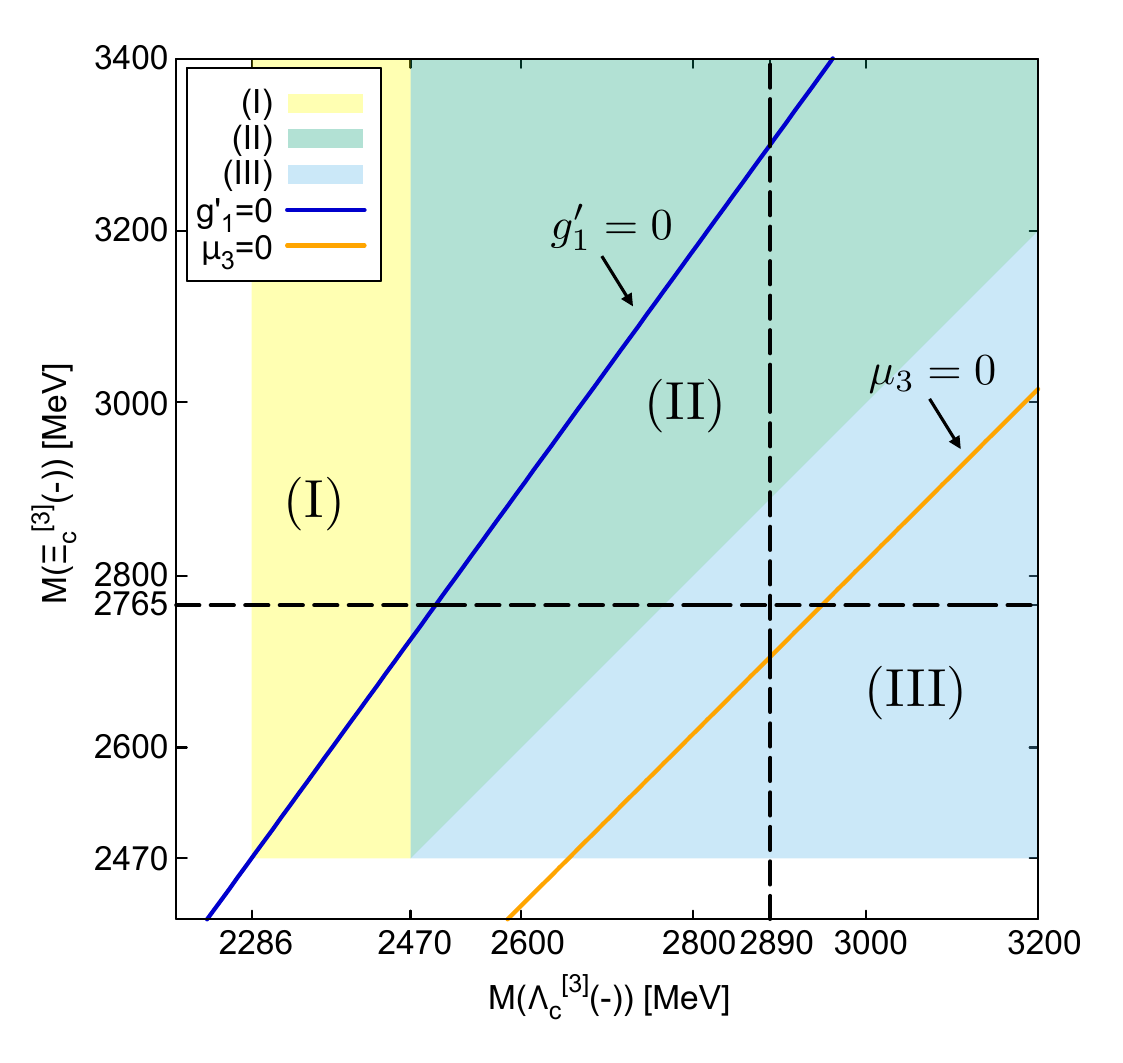}
\caption{Three types of the mass hierarchy listed in Eq.~(\ref{3QHierarchy}): (I) $M[{\Lambda}_c^{[\bm3]}(+)]<M[{\Lambda}_c^{[\bm3]}(-)]<M[{\Xi}_c^{[\bm3]}(+)]<M[{\Xi}_c^{[\bm3]}(-)]$, (II) $M[{\Lambda}_c^{[\bm3]}(+)]<M[{\Xi}_c^{[\bm3]}(+)]<M[{\Lambda}_c^{[\bm3]}(-)]<M[{\Xi}_c^{[\bm3]}(-)]$, and (III) $M[{\Lambda}_c^{[\bm3]}(+)]<M[{\Xi}_c^{[\bm3]}(+)]<M[{\Xi}_c^{[\bm3]}(-)]<M[{\Lambda}_c^{[\bm3]}(-)]$. The detail is explained in the text.
}
\label{fig:3QRegion}
\end{figure}

Taking into account those properties, the mass ordering of the negative-parity 3-quark SHBs is classified into the following three patterns:
\begin{eqnarray}
 M[{\Lambda}_c^{[\bm3]}(+)]<M[{\Lambda}_c^{[\bm3]}(-)]<M[{\Xi}_c^{[\bm3]}(+)]<M[{\Xi}_c^{[\bm3]}(-)]\, , \nonumber\\ 
 M[{\Lambda}_c^{[\bm3]}(+)]<M[{\Xi}_c^{[\bm3]}(+)]<M[{\Lambda}_c^{[\bm3]}(-)]<M[{\Xi}_c^{[\bm3]}(-)]\, , \nonumber\\
 M[{\Lambda}_c^{[\bm3]}(+)]<M[{\Xi}_c^{[\bm3]}(+)]<M[{\Xi}_c^{[\bm3]}(-)]<M[{\Lambda}_c^{[\bm3]}(-)]\, . \nonumber\\ \label{3QHierarchy}
\end{eqnarray}
In the first and second orderings, the negative-parity SHBs satisfy $M[{\Lambda}_c^{[\bm3]}(-)]<M[{\Xi}_c^{[\bm3]}(-)]$ similarly to the positive-parity ones, as naively expected from their flavor contents. For this reason, we call this mass ordering the normal mass hierarchy. In contrast, the third ordering in Eq.~(\ref{3QHierarchy}) indicates $M[{\Xi}_c^{[\bm3]}(-)]<M[{\Lambda}_c^{[\bm3]}(-)]$ which contradicts with the naive expectation, and this is referred to as the inverse mass hierarchy~\cite{Harada:2019udr}.

The three mass hierarchies~(\ref{3QHierarchy}) for $\Lambda_c^{[\bm 3]}(-)$ and $\Xi_c^{[\bm 3]}(-)$ are displayed in Fig.~\ref{fig:3QRegion}. In this figure, the colored regions (I), (II) and (III) correspond to the first, second and third hierarchies in Eq.~(\ref{3QHierarchy}), respectively. In Fig.~\ref{fig:3QRegion}, the mass hierarchy satisfied with $g_1'=0$ is denoted by the blue line, which always lies in the region of the normal mass hierarchy. That is, the inverse mass hierarchy for the negative-parity 3-quark SHBs does not manifest itself unless the $U(1)_A$ anomaly effects are present. The orange line with $\mu_3=0$ corresponds to the result in Ref.~\cite{Harada:2019udr}, which is included as a prominent example where the $U(1)_A$ anomaly effects are present. In fact, when $\mu_3=0$ one can prove the inverse mass hierarchy analytically as
\begin{eqnarray}
&&M[{\Lambda}_c^{[\bm3]}(-)]-M[{\Xi}_c^{[\bm3]}(-)] \nonumber\\
&& = M[{\Xi}_c^{[\bm3]}(+)]-M[{\Lambda}_c^{[\bm3]}(+)] >0\ ,
\end{eqnarray}
from Eqs.~(\ref{Mass3Q}) and~(\ref{GSInput}). The vertical and horizontal dashed lines represent a theoretical prediction of $M[{\Lambda}_c^{[\bm3]}(-)]=2890^*$ MeV from a quark model~\cite{Yoshida:2015tia} and that of $M[{\Xi}_c^{[\bm3]}(-)]=2765^*$ MeV from a diquark-heavy-quark potential model~\cite{Kim:2020imk}, respectively.\footnote{The asterisk in $2890^*$ is added to emphasize that the mass is a theoretical prediction. Throughout this article, we attach the asterisk (*) when referring to a theoretical prediction.} As seen from Fig.~\ref{fig:3QRegion}, a significant anomaly effect is necessary when we reproduce these theoretical predictions in our present approach. We note that lower limits of $M[{\Lambda}_c^{[\bm3]}(-)]$ and $M[{\Xi}_c^{[\bm3]}(-)]$ are constrained by Eq.~(\ref{3QBound}).

In what follows, we evaluate decay widths of the negative-parity SHBs induced by one-pseudoscalar-meson emissions in the absence of the mixing effects~(\ref{LMix}). Those coupling properties are read by taking fluctuations of the pseudoscalar mesons denoted by $P$ in addition to the VEVs~(\ref{VEV}) for the meson field $\Sigma$ as
\begin{eqnarray}
\Sigma \to \langle\Sigma\rangle + iP\ ,
\end{eqnarray}
with 
\begin{eqnarray}
P&=& \sqrt{2}  \nonumber\\
&\times& \begin{pmatrix}
\frac{\pi^0}{\sqrt{2}}+\frac{\eta_8}{\sqrt{6}}+\frac{\eta_1}{\sqrt{3}} & \pi^+ & K^+ \\
\pi^- & -\frac{\pi^0}{\sqrt{2}}+\frac{\eta_8}{\sqrt{6}}+\frac{\eta_1}{\sqrt{3}} & K^0 \\
K^- & \bar{K}^0 & -\frac{2\eta_8}{\sqrt{6}}+\frac{\eta_1}{\sqrt{3}}
\end{pmatrix}
\ . \nonumber\\ \label{NGBoson}
\end{eqnarray}
In Eq.~(\ref{NGBoson}), $\eta_1$ and $\eta_8$ are isospin-singlet pseudoscalar mesons belonging to flavor $SU(3)_{L+R}$ singlet and octet, respectively, which are not physical states due to a mixing between them. The physical states $\eta$ and $\eta'$ are defined by
\begin{eqnarray}
\begin{pmatrix}
\eta \\
\eta^{\prime}
\end{pmatrix}
=
\begin{pmatrix}
\cos\theta_P & -\sin\theta_P \\
\sin\theta_P & \cos\theta_P
\end{pmatrix}
\begin{pmatrix}
\eta_8 \\
\eta_1
\end{pmatrix}
\ ,
\end{eqnarray}
where the mixing angle $\theta_P$ is fixed to be $\theta_p=-11.3^\circ$ by the particle data group (PDG)~\cite{Workman:2022ynf}. 

\begin{figure*}[t]
\centering
\includegraphics*[scale=0.73]{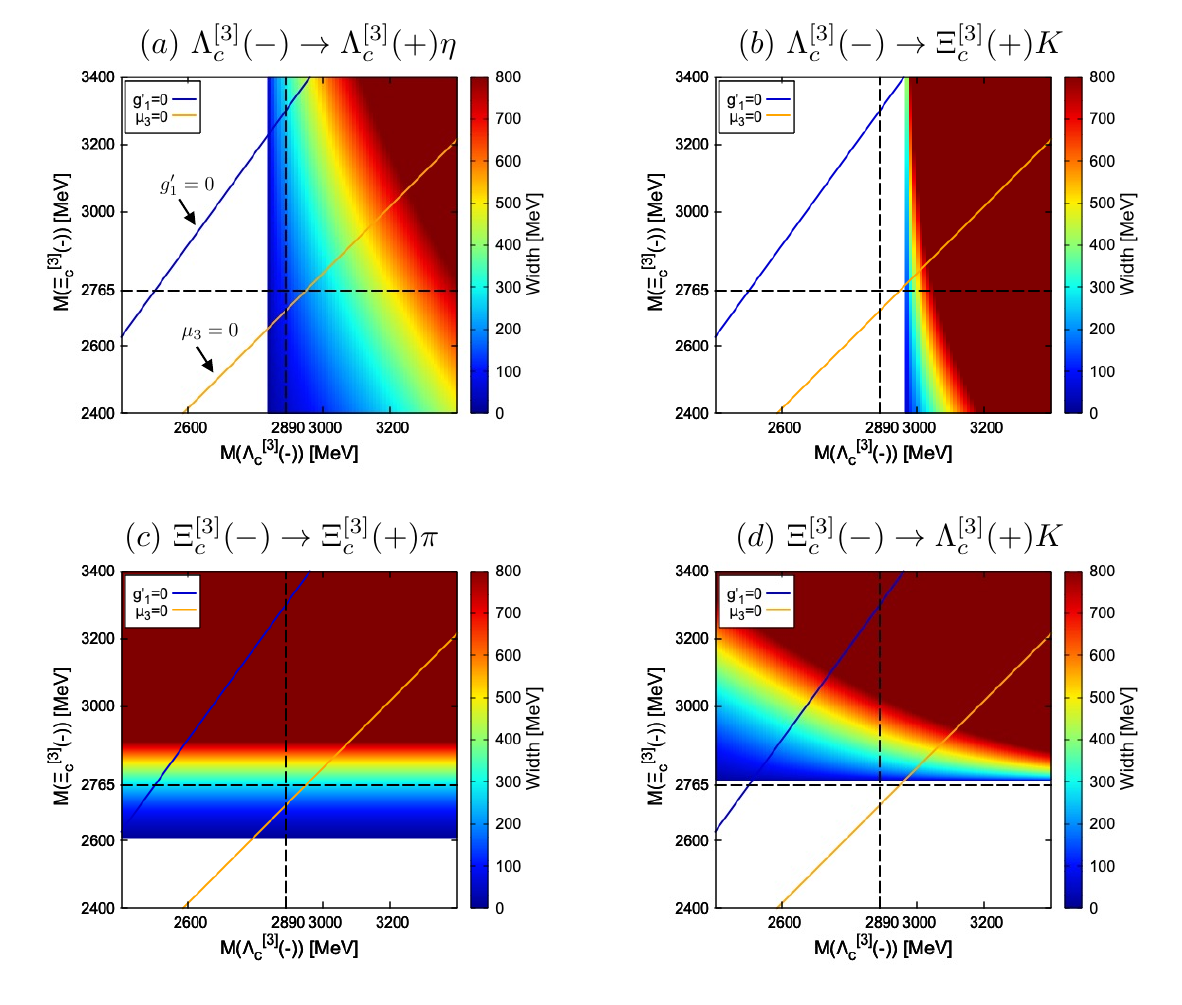}
  \caption{Partial decay widths of the negative-parity SHBs for arbitrary values of $M[\Xi_c^{[\bm3]}(-)]$ and $M[\Lambda_c^{[\bm3]}(-)]$. The blue and orange lines represent $g^{\prime}_1 = 0$ and $\mu_3 = 0$, respectively. The vertical and horizontal dashed lines are theoretical predictions of $M[{\Lambda}_c^{[\bm3]}(-)]=2890^*$ MeV and $M[{\Xi}_c^{[\bm3]}(-)]=2765^*$ MeV, respectively.}
\label{fig:3QDecay}
\end{figure*}

Having derived the coupling constant for decays of the one-pseudoscalar-meson emissions analytically, one can find that they are related to mass differences between the chiral partners as
\begin{eqnarray}
G_{\Xi^{[{\bm 3}]}_c(-)\Xi^{[{\bm 3}]}_c(+)\pi} = \frac{\Delta M(\Xi_c)}{2f_\pi}\ , \label{G3Q1} 
\end{eqnarray}
\begin{eqnarray}
G_{\Lambda^{[{\bm 3}]}_c(-)\Lambda^{[{\bm 3}]}_c(+)\eta} &=& \frac{\Delta M(\Lambda_c)+\Delta M(\Xi_c)}{\sqrt3 f_{\pi}(A+1)} \nonumber\\
&\times&\left(\cos \theta_P+\frac{\sin \theta_p}{\sqrt2}\right)\ , 
\label{G3Q2}
\end{eqnarray}
\begin{eqnarray}
G_{\Xi^{[{\bm 3}]}_c(-)\Lambda^{[{\bm 3}]}_c(+)K} =  \frac{\Delta M(\Lambda_c)+\Delta M(\Xi_c)}{\sqrt2 f_{\pi}(A+1)} \ , \label{G3Q3}
\end{eqnarray}
and
\begin{eqnarray}
G_{\Lambda^{[{\bm 3}]}_c(-)\Xi^{[{\bm 3}]}_c(+)K} =  \frac{\Delta M(\Lambda_c)+\Delta M(\Xi_c)}{\sqrt2 f_{\pi}(A+1)} \ ,\label{G3Q4}
\end{eqnarray}
with
\begin{eqnarray}
\Delta M(\Lambda_c) &\equiv&M[\Lambda_c^{[\bm3]}(-)]-M[\Lambda_c^{[\bm3]}(+)] \ ,\nonumber\\
\Delta M(\Xi_c) &\equiv& M[\Xi_c^{[\bm3]}(-)]-M[\Xi_c^{[\bm3]}(+)]\ .
\end{eqnarray}
Here, for instance, Eq.~(\ref{G3Q1}) stands for the coupling constant for a decay of $\Xi_c^{[\bm 3]}(-)\to\Xi_c^{[\bm 3]}(+)\pi$. The relations~(\ref{G3Q1}) -~(\ref{G3Q4}) are understood as extended-Goldberger-Treiman (GT) relations in our chiral model for the SHBs~\cite{Harada:2019udr,Kawakami:2020sxd}. In other words, the decay widths are solely determined by the masses of SHBs regardless of details of the model parameters, when the axial coupling is fixed to be unity as in the present linear sigma model. Among the relations, $G_{\Xi^{[{\bm 3}]}_c(-)\Xi^{[{\bm 3}]}_c(+)\pi}$ does not include the mass difference $\Delta M(\Lambda_c)$ since both the initial and final states are $\Xi_c$ baryons. Equation~(\ref{G3Q2}) indicates that the coupling $G_{\Lambda^{[{\bm 3}]}_c(-)\Lambda^{[{\bm 3}]}_c(+)\eta}$ is not only determined by $\Delta M(\Lambda_c)$ but also $\Delta M(\Xi_c)$ despite the absence of $\Xi_c$ in the reaction. Such a peculiar structure is induced by the anomaly effect which mixes all flavors. To see this we rewrite $G_{\Lambda^{[{\bm 3}]}_c(-)\Lambda^{[{\bm 3}]}_c(+)\eta}$ to
\begin{eqnarray}
G_{\Lambda^{[{\bm 3}]}_c(-)\Lambda^{[{\bm 3}]}_c(+)\eta} &=& \Bigg( \frac{\Delta M(\Lambda_c) }{\sqrt{3}f_\pi} - \frac{2}{\sqrt{3}}(A-1)g_1'\Bigg) \nonumber\\
&\times& \left(\cos \theta_P+\frac{\sin \theta_p}{\sqrt2}\right) \ . \label{GTLambda}
\end{eqnarray}
This equation indeed shows that the coupling constant is determined by only $\Delta M(\Lambda_c)$ in the absence of the anomaly effect: $g_1'=0$. Also, when $g_1'>0$, Eq.~(\ref{GTLambda}) indicates that the decay width of $\Lambda_c^{[3]}(-)$ is surppressed compared with a simple estimation obtained from the naive use of GT relation, $G = \Delta M/\sqrt{3} f_\pi$~\cite{Kawakami:2020sxd}.

Using the coupling constants in Eqs.~(\ref{G3Q1}) -~(\ref{G3Q4}), partial decay widths of the negative-parity SHBs for arbitrary values of $M[\Xi_c^{[\bm3]}(-)]$ and $M[\Lambda_c^{[\bm3]}(-)]$ are evaluated as displayed in Fig.~\ref{fig:3QDecay}. 
The four subfigures show, by the color map, the decay widths of (a) $\Lambda^{[{\bm 3}]}_c(-) \to \Lambda^{[{\bm 3}]}_c(+)\eta$, (b) $\Lambda^{[{\bm 3}]}_c(-) \to \Xi^{[{\bm 3}]}_c(+)K$, (c) $\Xi^{[{\bm 3}]}_c(-) \to \Xi^{[{\bm 3}]}_c(+)\pi$ and (d) $\Xi^{[{\bm 3}]}_c(-) \to \Lambda^{[{\bm 3}]}_c(+)K$. The blue line represents $g^{\prime}_1 = 0$ denoting the absence of the anomaly effects, and the orange one represents $\mu_3 = 0$ corresponding to the analysis in Ref.~\cite{Harada:2019udr}. The vertical and horizontal dashed lines are theoretical predictions of $M[{\Lambda}_c^{[\bm3]}(-)]=2890^*$ MeV~\cite{Yoshida:2015tia} and $M[{\Xi}_c^{[\bm3]}(-)]=2765^*$ MeV~\cite{Kim:2020imk}, respectively.

Figure~\ref{fig:3QDecay} indicates that the decay widths become large immediately when the thresholds open. Within the chiral model where relevant couplings are controlled by the extended GT relations, the decay widths are proportional to the square of mass differences between the chiral partners as seen from Eqs.~(\ref{G3Q1}) -~(\ref{G3Q4}). Moreover, $S$-wave decay rates are proportional to the momentum of the emitted pseudoscalar meson which are basically determined by the mass differences again. Hence, in total the decay widths are found to be proportional to the third power of the mass differences, which results in the rapid growth of the decay widths when the mass difference increases. 
We note that the decay width of $\Xi^{[{\bm 3}]}_c(-) \to \Xi^{[{\bm 3}]}_c(+)\pi$ is not affected by the mass of $M[\Lambda_c^{[{\bm 3}]}(-)]$ since the coupling is given by $\Delta M(\Xi_c)$ solely. We also note that, for $\Lambda^{[{\bm 3}]}_c(-) \to \Lambda^{[{\bm 3}]}_c(+)\eta$, there is a rather wide area for comparably small decay width, particularly in the region where the inverse mass hierarchy is realized, thanks to the $\eta$-$\eta'$ mixing as explained in Ref.~\cite{Kawakami:2020sxd}. We emphasize that there is no room to discuss such a broad detectable region unless the anomaly effects denoted by a nonzero value of $g_1'$ are present. When we take $M[{\Lambda}_c^{[\bm3]}(-)]=2890^*$ MeV and $M[{\Xi}_c^{[\bm3]}(-)]=2765^*$ MeV from the theoretical predictions, the resultant partial decay widths read $120$ MeV for $\Lambda^{[{\bm 3}]}_c(-) \to \Lambda^{[{\bm 3}]}_c(+)\eta$ and $264$ MeV for $\Xi^{[{\bm 3}]}_c(-) \to \Xi^{[{\bm 3}]}_c(+)\pi$, and the remaining two decay modes are closed.

As indicated in the PDG, $\Xi_c(2923)$ or $\Xi_c(2930)$ whose spin and parity are unknown can be candidates of $\Xi_c^{[{\bm 3}]}(-)$ in our present analysis~\cite{Workman:2022ynf}. Experimentally, the total decay widths of $\Xi_c(2923)$ and $\Xi_c(2930)$ are known to be $\Gamma^{\rm tot}_{\Xi_c(2930)^+} = 15\pm9$ MeV [$\Gamma^{\rm tot}_{\Xi_c(2930)^0} = 10.2\pm1.4$ MeV], whereas our prediction yields significantly larger decay widths as seen from Fig.~\ref{fig:3QDecay} (c) and (d). For this reason we conclude that the observed $\Xi_c(2923)$ or $\Xi_c(2930)$ is not identified as a 3-quark $\Xi_c^{[{\bm 3}]}(-)$. In Sec.~\ref{sec:35Mixing}, we show that $\Xi_c(2923)$ or $\Xi_c(2930)$ would be identified with a 5-quark dominant SHB where only a small fraction of the 3-quark one enters.

\section{Analysis of 5-quark SHBs}
\label{sec:5-quark}

In this section, we investigate mass spectrum and decay widths of the 5-quark SHBs from Eq.~(\ref{L5q}). Similarly to the analysis in Sec.~\ref{sec:3-quark}, here we switch off mixings from the 3-quark SHBs by omitting Eq.~(\ref{LMix}) so as to gain clear insight into properties of the 5-quark SHBs from chiral symmetry and the $U(1)_A$ anomaly.

Parity eigenstates, i.e., mass eigenstates of the 5-quark SHBs are obtained by linear combinations of $B_R'$ and $B_L'$ as
\begin{eqnarray}
 B'_{\pm}=\frac{1}{\sqrt{2}}(B'_R\mp B'_L)\ , \label{Parity5q}
\end{eqnarray}
and from Eq.~(\ref{L5q}) the corresponding mass eigenvalues read 
\begin{eqnarray}
M[{\Lambda}_c^{[\bm5]}(\pm)] &=& m_B + \mu_2+A^2\mu_4 \nonumber\\
&& \pm Af_{\pi}\big[A(g_2+g_3)+g_2'\big]\ ,\nonumber\\
M[{\Xi}_c^{[\bm5]}(\pm)]&=&m_B + \mu_2+\mu_4\pm f_{\pi}\big[A(g_2+g_3)+g_2'\big] \ . \nonumber\\
 \label{Mass5Q}
\end{eqnarray}
The notation in these equations follows Eq.~(\ref{Mass3Q}), and the quark contents are $uds\bar{u}c$ ($uds\bar{d}c$) in ${\Xi}_c^{[\bm5]}(\pm)$ and $uds\bar{s}c$ in ${\Lambda}_c^{[\bm5]}(\pm)$. Equation~(\ref{Mass5Q}) indicates that the mass formulas for ${\Lambda}_c^{[\bm5]}(\pm)$ and ${\Xi}_c^{[\bm5]}(\pm)$ share a common piece of $A(g_2+g_3)+g_2'$ in the last term, and thereby, we can absorb the three parameters $g_2$, $g_3$ and $g_2'$ into a single parameter $h$ as
\begin{eqnarray}
M({\Lambda}_c^{[\bm5]}(\pm)) &=&m_B + \mu_2+A^2\mu_4\pm Af_{\pi}h\ , \nonumber\\
M({\Xi}_c^{[\bm5]}(\pm)) &=& m_B + \mu_2+\mu_4\pm f_{\pi}h\ .
 \label{Mass5Q2}
\end{eqnarray}
For this reason, now the number of free parameters is three: $\mu_2$, $\mu_4$ and $h$. From Eq.~(\ref{Mass5Q2}), one can conclude that the leading contributions from the $U(1)_A$ anomaly incorporated by the $g_2'$ term do not affect the mass formula, which is distinct from the case of 3-quark SHBs where the anomalous term plays a significant role for the mass hierarchy. Accordingly, the $U(1)_A$ anomaly does not contribute to the decay widths stemming from one-pseudoscalar-meson emissions due to the extended-GT relation for the 5-quark SHBs. Another characteristic feature is the influence of the violation of $SU(3)_{L+R}$ flavor symmetry, that is, $A^2$ appears as a coefficient of $\mu_4$ for $M[\Lambda_c^{[{\bm 5}]}(\pm)]$, which is again distinct from the case of 3-quark SHBs where $A^2\mu_3$ appears for $M[\Xi_c^{[{\bm 3}]}(\pm)]$. Such a noteworthy feature is understood by the quark-line diagram in Fig.~\ref{fig:L5}. In fact, as seen from the diagram for the $\mu_4$ term, when we focus on the $\Lambda_c^{[{\bm 5}]}(\pm)$ baryons composed of $uds\bar{s}c$, the two $\Sigma^{(\dagger)}$ couples with an $\bar{s}$ line which generates $A^2$ contributions in the mass formula.


As for the 5-quark SHBs, we identify $\Lambda_c^{[{\bm 5}]}(+)$ and $\Xi_c^{[{\bm 5}]}(+)$ with the experimentally observed Roper-like states, $\Lambda_c(2765)$ and $\Xi_c(2970)$, respectively. Then~\cite{Workman:2022ynf}
\begin{eqnarray}
M[{\Lambda}_c^{[\bm5]}(+)] &=& 2765\, {\rm MeV} \ , \nonumber\\
M[{\Xi}_c^{[\bm5]}(+)] &=& 2967\, {\rm MeV} \ . \label{RoperMass}
\end{eqnarray}
Here, the 5-quark SHBs include one antiquark whose intrinsic parity is $-1$, so we expect that $\Lambda_c^{[{\bm 5}]}(-)$ and $\Xi_c^{[{\bm 5}]}(-)$ are regarded as the ground states while $\Lambda_c^{[{\bm 5}]}(+)$ and $\Xi_c^{[{\bm 5}]}(+)$ are the orbitally excited states. Hence, one can naturally assume the following mass hierarchies:
\begin{eqnarray} 
M[{\Lambda}_c^{[\bm5]}(-)] &<& M[{\Lambda}_c^{[\bm5]}(+)] \ , \nonumber\\
M[{\Xi}_c^{[\bm5]}(-)] &<& M[{\Xi}_c^{[\bm5]}(+)]  \ .  \label{MassRelation5q}
\end{eqnarray}
Other constraints for the mass hierarchy are obtained from decay widths of $\Lambda_c(2765)$ and $\Xi_c(2970)$. Experimentally the total decay widths of these SHBs are known to be $\Gamma^{\rm tot}_{\Lambda_c(2765)} \approx 50$ MeV and $\Gamma^{\rm tot}_{\Xi_c(2970)} \approx 20.9$ MeV~\cite{Workman:2022ynf}. Thus, these values are regarded as the upper limits of the partial decay widths due to one-pseudoscalar-meson emissions:
\begin{eqnarray}
&&\Gamma(\Xi_c(2970)\to\Xi_c^{[\bm5]}(-)\pi) + \Gamma(\Xi_c(2970)\to\Lambda_c^{[\bm5]}(-)K) \nonumber\\
 && \hspace{5cm}  \lesssim 20.9\,  {\rm MeV}\ , \nonumber\\
&& \Gamma(\Lambda_c(2765)\to\Xi_c^{[\bm5]}(-)K) + \Gamma(\Lambda_c(2765) \to \Lambda_c^{[\bm5]}(-)\eta) \nonumber\\
&&  \hspace{5cm}  \lesssim 50\, {\rm MeV}\ .  \label{RoperDecay}
\end{eqnarray}
Similarly to decays of the 3-quark SHBs whose couplings are determined by mass differences of the chiral partners as in Eqs.~(\ref{G3Q1}) -~(\ref{G3Q4}), decay widths of the 5-quark SHBs shown in Eq.~(\ref{RoperDecay}) are also expressed by the mass differences regardless of details of the model. In other words, Eq.~(\ref{RoperDecay}) enables us to get constraints on the masses of $\Lambda_c^{[\bm5]}(-)$ and $\Xi_c^{[\bm5]}(-)$ directly, which yields
\begin{eqnarray}
2551\, {\rm MeV} &\lesssim& M[{\Lambda}_c^{[\bm5]}(-)]\ , \nonumber\\
2811\, {\rm MeV} &\lesssim& M[{\Xi}_c^{[\bm5]}(-)]\ . \label{5qConstraint}
\end{eqnarray}
Notably, under these constraints the decay modes, $\Xi_c(2970)\to\Lambda_c^{[\bm5]}(-)K$, $\Lambda_c(2765)\to\Xi_c^{[\bm5]}(-)K$ and $\Lambda_c(2765) \to \Lambda_c^{[\bm5]}(-)\eta$, are closed and only $\Xi_c(2970)\to\Xi_c^{[\bm5]}(-)\pi$ is allowed, resulting in the disappearance of decays of $\Lambda_c(2765)$ induced by the one-pseudoscalar-meson emission. The main decay modes of $\Lambda_c(2765)$ are sequential decays emitting two pions via $\Sigma_c$ resonances~\cite{Arifi:2020ezz,Arifi:2020yfp,Suenaga:2022ajn}, which are not treated in our present model.\footnote{We have employed the PDG value of $\Gamma^{\rm tot}_{\Lambda_c(2765)} \approx 50$ MeV to find the constraints~(\ref{5qConstraint}) although, e.g., the Belle collaboration reported a larger value of $\Gamma^{\rm tot}_{\Lambda_c(2765)} = 73\pm5$ MeV~\cite{Belle:2006xni}. However, the constraints in Eq.~(\ref{5qConstraint}) are not significantly affected by variations of $\Gamma^{\rm tot}_{\Lambda_c(2765)}$ which are dominated by the sequential two-pion emission decays.}
Combining Eqs.~(\ref{MassRelation5q}) and~(\ref{5qConstraint}), the mass hierarchy of the 5-quark SHBs is uniquely determined to be
\begin{eqnarray}
M[{\Lambda}_c^{[\bm5]}(-)]< M[{\Lambda}_c^{[\bm5]}(+)] < M[{\Xi}_c^{[\bm5]}(-)] < M[{\Xi}_c^{[\bm5]}(+)] \ . \nonumber\\ \label{5qHierarchy}
\end{eqnarray}
This mass ordering may not be intuitive since $\Lambda^{[\bm 5]}_c(\pm)$ is heavier than $\Xi_c^{[\bm 5]}(\pm)$ despite their quark contents: $\Lambda_c^{[\bm 5]}(\pm) \sim uds\bar{s}c $ and $\Xi_c^{[\bm 5]}(\pm) \sim uds\bar{u}c$ ($uds\bar{d}$c). A possible scenario to obtain such unnatural mass ordering is discussed in Appendix.~\ref{sec:Interpretation}.

The experimentally observed $\Xi_c(2923)$ or $\Xi_c(2930)$ are expected to be candidates of $\Xi_c^{[{\bm 5}]}(-)$, since the mass of $\Xi_c(2923)$ or $\Xi_c(2930)$ satisfies the inequality in Eq.~(\ref{5qHierarchy}).\footnote{The masses of $\Xi_c(2923)$ and $\Xi_c(2930)$ read $M[\Xi_c(2923)]\approx 2923$ MeV and $M[\Xi_c(2930)]\approx2939$ MeV, respectively~\cite{Workman:2022ynf}.} In Sec.~\ref{sec:35Mixing}, indeed, we show that $\Xi_c(2923)$ or $\Xi_c(2930)$ can be identified with the negative-parity 5-quark dominant $\Xi_c$ from its decay properties.

 As for $\Lambda_c^{[{\bm 5}]}(-)$, one can see that $\Lambda_c^{[{\bm 5}]}(-)$ does not exhibit strong decays from the constraint on the mass in Eq.~(\ref{5qHierarchy}), as long as the dynamics is governed by exact HQSS. Such stable behavior holds even after introducing mixings with the 3-quark SHBs. Its possible strong decay induced by a violation of HQSS is discussed in Sec.~\ref{sec:DecayLambdaC}. We note that, when we identify $\Xi_c^{[{\bm 5}]}(-)$ with $\Xi_c(2923)$ or $\Xi_c(2930)$, the mass of $\Lambda_c^{[{\bm 5}]}(-)$ reads $M[{\Lambda}_c^{[\bm5]}(-)] = 2704$ MeV or $M[{\Lambda}_c^{[\bm5]}(-)] = 2726$ MeV.

\section{Analysis with mixings between 3-quark and 5-quark SHBs}
\label{sec:35Mixing}

From the analysis in Secs.~\ref{sec:3-quark} and~\ref{sec:5-quark}, we have learned that the $U(1)_A$ axial anomaly can lead to the inverse mass hierarchy for the negative-parity 3-quark SHBs while it does not affect the mass spectrum of the 5-quark SHBs. In this section, we generalize the discussion by including mixings between the 3-quark and 5-quark SHBs to delineate the realistic spectrum of the SHBs, and present predictions based on our model.

\subsection{Mass formula}
\label{sec:MixMass}

Here, we present the mass formula of the SHBs with mixings between the $3$-quark and $5$-quark components.

In the presence of the mixings, mass eigenstates take the form of~\cite{Suenaga:2021qri}
\begin{align}
\left(
\begin{array}{c}
B_{\pm,i}^L \\
B_{\pm,i}^H \\
\end{array}
\right) = \left(
\begin{array}{cc}
\cos\theta_{B_{\pm,i}} & \sin\theta_{B_{\pm,i}} \\
-\sin\theta_{B_{\pm,i}} & \cos\theta_{B_{\pm,i}} \\
\end{array}
\right)\left(
\begin{array}{c}
B_{\pm,i} \\
B_{\pm,i}' \\
\end{array}
\right)\ , \label{Mixing}
\end{align}
where the mixing angles satisfy $\tan2\theta_{B_{\pm,i}} =(2\tilde{m}_{\pm,i})/(m^{[{\bm 2}]}_{\pm,i}-m^{[{\bm 4}]}_{\pm,i})$, and the corresponding mass eigenvalues read
\begin{eqnarray}
 M(B^{H/L}_{+,i}) &=& m_B + \frac{1}{2}\Bigg[m^{[\bm 2]}_{+,i}+m^{[\bm 4]}_{+,i} \nonumber\\
 && \pm\sqrt{{\left(m^{[\bm 2]}_{+,i}-m^{[\bm 4]}_{+,i}\right)}^{2}+4\tilde{m}^2_{+,i}}\  \Bigg]\ , \nonumber\\
M(B^{H/L}_{-,i}) &=& m_B + \frac{1}{2}\Bigg[m^{[\bm 2]}_{-,i}+m^{[\bm 4]}_{-,i} \nonumber\\
&& \pm\sqrt{{\left(m^{[\bm 2]}_{-,i}-m^{[\bm 4]}_{-,i}\right)}^{2}+4\tilde{m}^2_{-,i}}\  \Bigg]\ , \label{MassMix}
\end{eqnarray}
with
\begin{eqnarray}
m^{[\bm 2]}_{\pm,i=1,2}&=&\mu_1+A^2\mu_3\mp f_{\pi}(Ag_1+g_1^{\prime}),  \nonumber\\
    m^{[\bm 2]}_{\pm,i=3}&=&\mu_1+\mu_3\mp f_{\pi}(g_1+Ag_1^{\prime})\ , \nonumber\\
    m^{[\bm 4]}_{\pm,i=1,2}&=&\mu_2+\mu_4\pm f_{\pi}h \ , \nonumber\\
    m^{[\bm 4]}_{\pm,i=3}&=&\mu_2+A^2\mu_4\pm Af_{\pi}h \ , \nonumber\\
    \tilde{m}_{\pm,i=1,2}&=&\mu_1^{\prime}\mp f_{\pi}g_4\ , \nonumber\\
    \tilde{m}_{\pm,i=3}&=&\mu_1^{\prime}\mp Af_{\pi}g_4\ . \label{EachMass}
\end{eqnarray}
In Eqs.~(\ref{Mixing}) and~(\ref{MassMix}) the subscripts ``$\pm$'' and ``$i$'' in the $B_{\pm,i}^{H/L}$ stand for the parity and flavor indices, respectively. Besides, the superscript $H$ ($L$) represents the higher (lower) mass eigenstate corresponding to the plus (minus) sign in front of the square root in the right-hand side (RHS) of Eq.~(\ref{MassMix}). As for Eq.~(\ref{EachMass}), $\tilde{m}_{\pm,i}$ is responsible for the mixings, and $m_{\pm ,i}^{[{\bm 2}]}$ and $m_{\pm ,i}^{[{\bm 4}]}$ correspond to the masses of the pure diquarks ($qq$) and the tetra-diquarks ($qq\bar{q}q$), respectively. We note that masses~(\ref{MassMix}) satisfy 
\begin{eqnarray}
&& \sum_{p=\pm, n=L,H} M(B_{p,i=1,2}^{H/L}) - \sum_{p=\pm, n=L,H}  M(B_{p,i=3}^{H/L})\nonumber\\
&& = 2(A^2-1)(\mu_3-\mu_4)\ ,
\end{eqnarray}
which can be understood as a generalization of the simple mass formula found in Ref.~\cite{Suenaga:2021qri}:
\begin{eqnarray}
\sum_{p=\pm, n=L,H} M(B_{p,i=1,2}^{H/L}) = \sum_{p=\pm, n=L,H}  M(B_{p,i=3}^{H/L})\ .
\end{eqnarray}
The $\mu_3$ term produces differences between the parity-averaged masses of $\Lambda_c^{[{\bm 3}]}$ and $\Xi_c^{[{\bm 3}]}$, and so does the $\mu_4$ one for $\Lambda_c^{[{\bm 5}]}$ and $\Xi_c^{[{\bm 5}]}$, as seen from Eqs.~(\ref{Mass3Q}) and~(\ref{Mass5Q}). Such effects are generated by ${\cal O}(\Sigma^2)$ and were not incorporated in Ref.~\cite{Suenaga:2021qri}.

In what follows, similarly to the analysis in Secs.~\ref{sec:3-quark} and~\ref{sec:5-quark}, we employ the notation of $M[\Lambda_c^{H/L}(\pm)]$ and $M[\Xi_c^{H/L}(\pm)] $ to refer to the corresponding masses as follows:
\begin{eqnarray}
M[\Lambda_c^{H/L}(\pm)] &\equiv& M(B_{\pm,i=3}^{H/L})\ , \nonumber\\
M[\Xi_c^{H/L}(\pm)] &\equiv& M(B_{\pm,i=1,2}^{H/L})\ .
\end{eqnarray}

\subsection{Without anomaly effects}
\label{sec:MixNoAnomaly}

In this subsection, toward a clear understanding of the mixing effects on the mass and decay properties of the negative-parity SHBs, we proceed with the investigation without the $U(1)_A$ anomaly effects.

\begin{table}[t]
\begin{center}
  \begin{tabular}{cc}  \hline\hline
$M[\Lambda_c^{L}(+)] = 2286$ MeV & $M[{\Xi}_c^{L}(+)] = 2470$ MeV \\ 
$M[\Lambda_c^{H}(+)] = 2765$  MeV & $M[{\Xi}_c^{H}(+)] = 2967$ MeV  \\ \hline
$M[\Lambda_c(-)] = 2890^*$ MeV & $M[\Xi_c(-)] = 2939$ MeV \\ \hline \hline
 \end{tabular}
\caption{Input masses for the analysis in Sec.~\ref{sec:MixNoAnomaly}. For the negative-parity SHBs the mass orderings are obscure so that the superscript $H$ or $L$ is not attached.}
\label{tab:Parameter1}
\end{center}
\end{table}

In the absence of the anomaly effects, there are seven model parameters to be fixed: $\mu_1$, $\mu_2$, $\mu_3$, $\mu_4$, $g_1$, $h= A (g_2 + g_3)$ and $g_4$. Four of them are fixed from the masses of positive-parity SHBs, where the ground-state and the Roper-like SHBs are assumed to be 3-quark and 5-quark dominant, respectively. For this reason we assume $m_{+,i}^{[{\bm 2}]}<m_{+,i}^{[{\bm 4}]}$.
Besides, as a typical value for the mass of negative-parity $\Lambda_c$, we take the quark-model prediction of $M[\Lambda_c(-)] = 2890^*$ MeV
for another input. Furthermore, we employ the mass of $\Xi_c(2930)$ as an input. As explained at the end of Sec.~\ref{sec:3-quark}, $\Xi_c(2930)$ cannot be identified with the 3-quark dominant SHB from its decay width, and thus we suppose $\Xi_c(2930)$ is 5-quark dominant. The input masses are summarized in Table.~\ref{tab:Parameter1}.

\begin{figure}[t]
\centering
\hspace*{-0.5cm} 
\includegraphics*[scale=0.42]{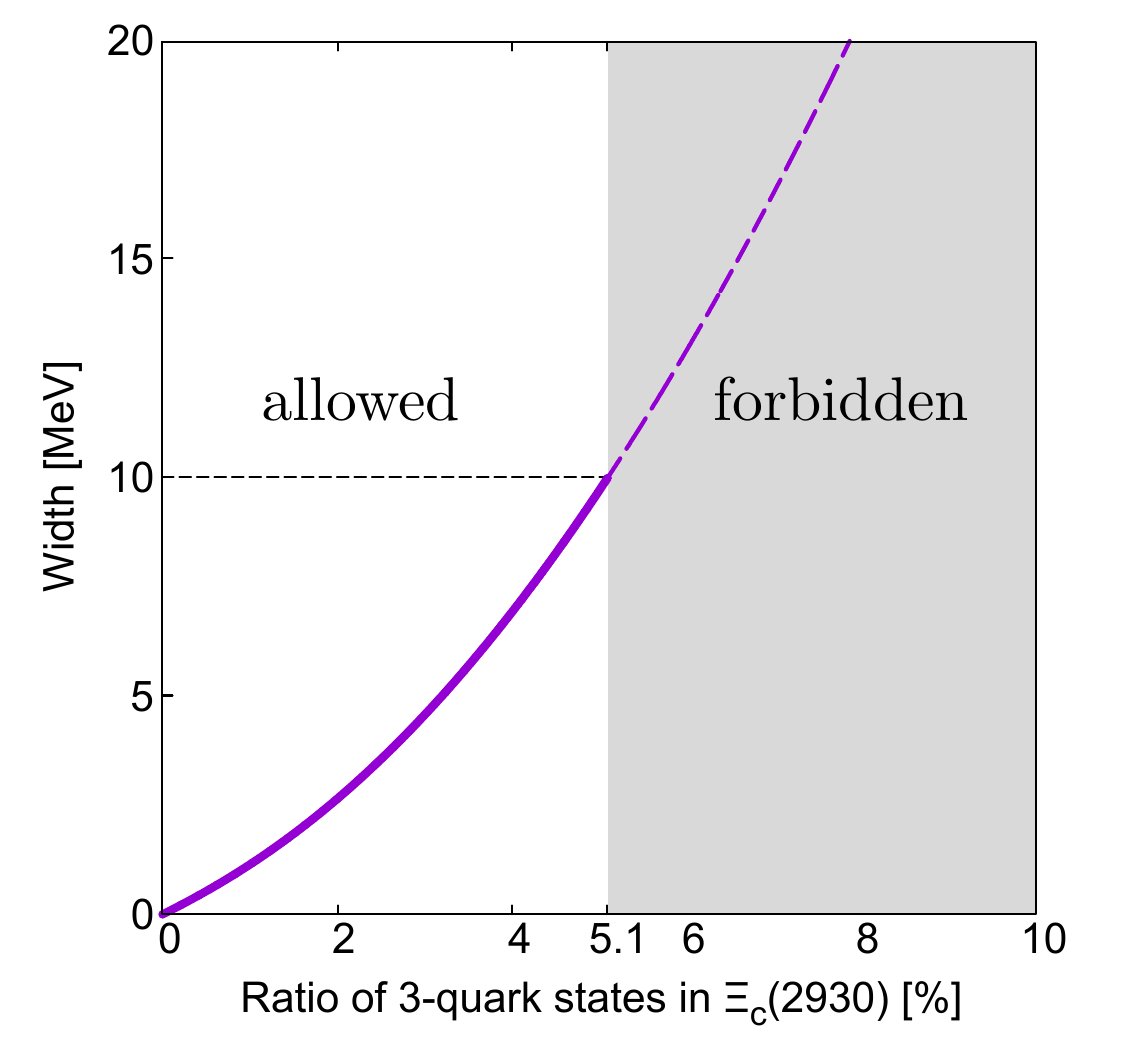}
\caption{Decay width of $\Xi_c(2930)$ as a function of the ratio of 3-quark states in $\Xi_c(2930)$, where the horizontal axis is defined by $100\times(\cos\theta_{B_{-,i=1,2}})^2$. The decay modes which can be treated in our present framework are $\Xi_c(2930)\to \Xi_c(2470)\pi$ and $\Xi_c(2930)\to \Lambda_c(2286)K$. Typically only the colorless area is allowed from the experimental data of decays of $\Xi_c(2930)$.}
\label{fig:DecayXiRatio}
\end{figure}

\begin{figure}[t]
\centering
\hspace{-0.5cm}
\includegraphics*[scale=0.46]{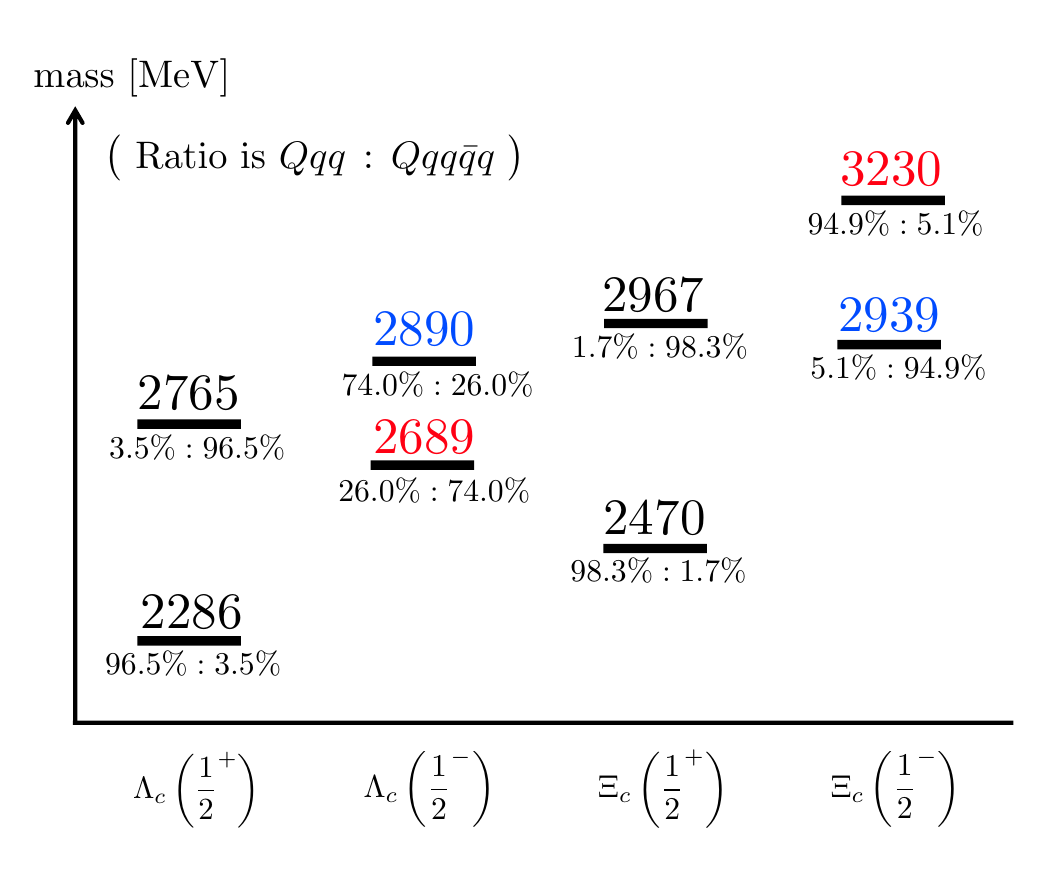}
  \caption{Mass spectrum of all SHBs treated in our present model with the parameter set~(\ref{PSet1}). The details are provided in the text.}
\label{fig:Spectrum1}
\end{figure}


Now, we can work with only one parameter. Choosing the mixing angle $\theta_{B_-,i=1,2}$ as the last parameter, for instance we can examine the decay width of $\Xi_c(2930)$ as a function of the ratio of 3-quark states in $\Xi_c(2930)$, as depicted in Fig.~\ref{fig:DecayXiRatio}. In this figure the horizontal axis is defined by $100\times(\cos\theta_{B_{-,i=1,2}})^2$. The decay modes which can be treated in our present framework are $\Xi_c(2930)\to \Xi_c(2470)\pi$ and $\Xi_c(2930)\to \Lambda_c(2286)K$, where $\Xi_c(2470)$ and $\Lambda_c(2286)$ are  reduced to pure 3-quark SHBs when the mixings are switched off. Hence, the decay width vanishes when the 3-quark component in $\Xi_c(2930)$ is zero due to the orthogonality of the initial and final states, corresponding to the consideration in Sec.~\ref{sec:5-quark}. Then, the width begins to grow as the ratio increases through the small overlaps. The PDG reads $\Gamma^{\rm tot}_{\Xi_c(2930)^+} = 15\pm9$ MeV [$\Gamma^{\rm tot}_{\Xi_c(2930)^0} = 10.2\pm1.4$ MeV], so that typically the ratio is allowed to be less than $\sim 5.1$ \% from which the width becomes $\sim 10$ MeV, as denoted by the the colorless area in Fig.~\ref{fig:DecayXiRatio}.

When we fix the last parameter at which the mixing of 3-quark states in $\Xi_c(2930)$ is $5.1$\%, all the seven parameters are determined to be
\begin{eqnarray}
&& \mu_1=-518\, {\rm MeV}\ , \ \ \mu_2=413\, {\rm MeV}\ , \ \ \mu_3=309\, {\rm MeV}\  , \nonumber\\
&& \mu_4 = -236\, {\rm MeV}\ , \ \ g_1=2.88\ , \ \ g_4 =-0.687\ , \nonumber\\
&&  h=0.0259\ .
\label{PSet1}
\end{eqnarray}
We note that we have taken $m_B=2780$ MeV in obtaining parameters~(\ref{PSet1}) such that $m_B$ coincides with the averaged mass of all eight SHBs. We also note that the dimensionless parameter $h$ originating from the ${\cal O}(\Sigma^4)$ contributions is indeed suppressed compared to $g_1$ and $g_4$ terms. With the parameter set~(\ref{PSet1}), the mass spectrum of all SHBs treated in our present model is obtained as displayed in Fig.~\ref{fig:Spectrum1}. In this figure, the mass values indicated by black and blue colors are inputs for the positive-parity SHBs and negative-parity SHBs, respectively, whereas the red values are outputs (see Table~\ref{tab:Parameter1}). The percentage below the mass values denotes the ratio of 3-quark and 5-quark states: $Qqq$ and $Qqq\bar{q}q$. The figure indicates that, for the positive-parity sector, the Roper-like SHBs are mostly 5-quark states while the ground-state ones are mostly 3-quark states. Meanwhile, for the negative-parity sector the tendency is opposite; the higher-mass SHBs are 3-quark dominant while the lower-mass ones are 5-quark dominant. Such a characteristic result follows our intuitive assumption for the positive-parity SHBs $m_{+,i}^{[{\bm 2}]}<m_{+,i}^{[{\bm 4}]}$ and a comparably small decay width of $\Xi_c(2930)$. 

Here, we discuss properties of $\Lambda_c^L(-)$ and $\Xi_c^H(-)$ which are our outputs. As for $\Lambda_c^L(-)$, the mass reads $M[\Lambda_c^L(-)]=2689$ MeV which is smaller than the result without the mixings estimated at the end of Sec.~\ref{sec:5-quark}: $M[\Lambda_c^{[{\bm 5}]}(-)]=2726$ MeV. Such a mass reduction is understood by a level repulsion owing to the mixing with the 3-quark state. In fact, the mass lies in a range of 
\begin{eqnarray}
2689\, {\rm MeV} < M[\Lambda_c^L(-)] < 2726\, {\rm MeV}\ , \label{LambdaCOut}
\end{eqnarray}
corresponding to the allowed region in Fig.~\ref{fig:DecayXiRatio}. From this consideration we can find that the $\Lambda_c(2890^*)$ must be 3-quark dominant, which is consistent with the fact that $\Lambda_c(2890^*)$ is indeed predicted as a $\rho$-mode excitation by the quark model including only three quarks~\cite{Yoshida:2015tia}. Besides, $\Lambda_c^L(-)$ becomes stable within our present approach where $SU(2)_h$ HQSS is exact. In fact, $\Lambda_c^L(-)$ can decay into $\Sigma_c\pi$ only when we include a violation of $SU(2)_h$ HQSS as discussed in Sec.~\ref{sec:DecayLambdaC}, leading to the width of order a few MeV.\footnote{The analysis in Sec.~\ref{sec:DecayLambdaC} is done in the presence of the $U(1)_A$ axial anomaly effect, but the resultant decay width of $\Lambda_c^L(-)$ is qualitatively the same as the one without the anomaly.} Therefore, we conclude that possible existence of such a very narrow $\Lambda_c(-)$ in the mass region given by Eq.~(\ref{LambdaCOut}) will be a challenge to experiment as a good test of our description based on mixings between the 3-quark and 5-quark SHBs. 

On the other hand, $\Xi_c^H(-)$ can decay into, e.g., $\Xi_c^L(+)\pi$ easily due to its large mass of $M[\Xi_c^H(-)] = 3230$ MeV, as analogously understood from Fig.~\ref{fig:3QDecay}. We note that
\begin{eqnarray}
3230\, {\rm MeV} < M[\Xi_{c}^H(-)] < 3301\, {\rm MeV}\ ,
 \end{eqnarray}
corresponding to the allowed area in Fig.~\ref{fig:DecayXiRatio}, which is always larger than $\Xi_c(2930)$. Accordingly, the decay width of $\Xi_c^H(-)$ is expected to be always catastrophically broad. Qualitatively a similar argument follows when we assign $\Xi_c(2923)$ to $\Xi_c^{L}(-)$.

Our demonstration in this subsection implies that the normal mass hierarchy for the negative-parity 3-quark dominant SHBs remains satisfied even in the presence of the mixings: $M[\Lambda_c^H(-)] < M[\Xi_c^H(-)]$, as displayed in Fig.~\ref{fig:Spectrum1}, similarly to our previous analysis without anomaly effects in Sec.~\ref{sec:3-quark}. This ordering does not change as long as we employ the mass and small decay width of $\Xi_c(2930)$ in addition to the mass of $\Lambda_c(2890^*)$ as inputs. On the other hand, in Ref.~\cite{Kim:2020imk} a 3-quark (dominant) $\Xi_c$  was predicted at 2765 MeV based on the diquark-heavy-quark potential approach. If we take this value as an input, then the scenario is drastically different from the spectrum in Fig.~\ref{fig:Spectrum1}, since in this case the inverse mass hierarchy for the 3-quark (dominant) SHBs emerges: $M[\Lambda_c(2890^*)]>M[\Xi_c(2765^*)]$, and indeed one can show that there is no solution unless the anomaly effects enter. Hence, in Sec.~\ref{sec:MixWithAnomaly} we demonstrate the roles of the $U(1)_A$ anomaly effects with the mixings by taking $M[\Xi_c(2765^*)]$ as another input.

\begin{figure}[t]
\centering
\hspace*{-0.5cm}
\vspace*{-0.6cm}
\includegraphics*[scale=0.44]{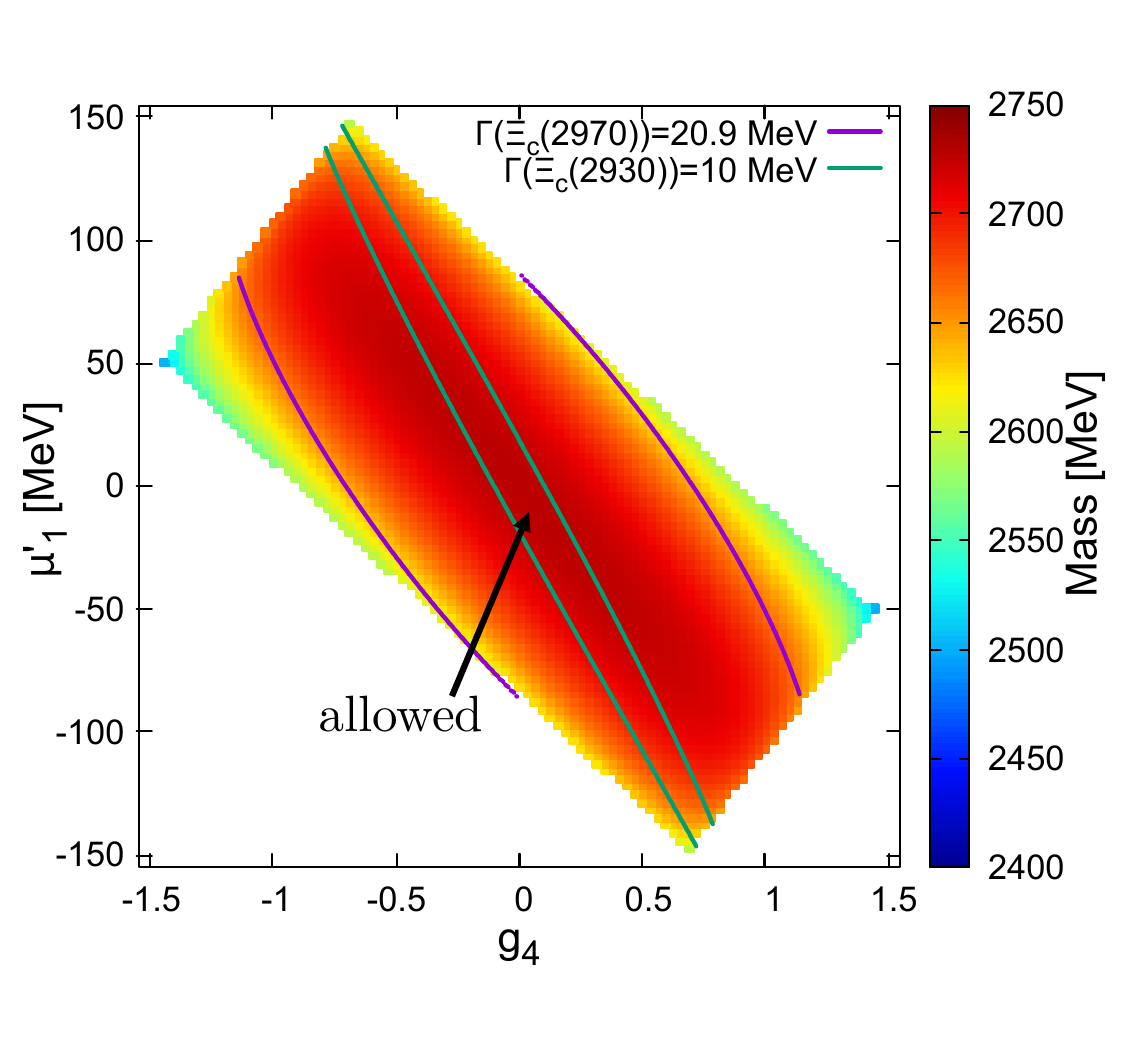}
  \caption{Mass of $\Lambda_c^L(-)$ in $\mu_1'$ - $g_4$ plane. The purple and green lines represent boundaries constrained from the decay widths of $\Xi_c(2970)$ and $\Xi_c(2930)$, respectively.}
\label{fig:LambdaMassAnomaly}
\end{figure}

\begin{table}[t]
\begin{center}
  \begin{tabular}{cc}  \hline\hline
$M(\Lambda_c^{L}(+)) = 2286$ MeV & $M({\Xi}_c^{L}(+)) = 2470$ MeV \\ 
$M(\Lambda_c^{H}(+)) = 2765$  MeV & $M({\Xi}_c^{H}(+)) = 2967$ MeV  \\ \hline
output &  $M(\Xi^L_c(-)) = 2765^*$ MeV \\ 
$M(\Lambda_c^H(-)) = 2890^*$ MeV & $M(\Xi^H_c(-)) = 2939$ MeV \\ \hline \hline
 \end{tabular}
\caption{Input masses for the analysis in Sec.~\ref{sec:MixWithAnomaly}.}
\label{tab:Parameter2}
\end{center}
\end{table}

\subsection{With anomaly effects}
\label{sec:MixWithAnomaly}

In this subsection, we include the $U(1)_A$ anomaly effects together with the mixing of 3-quark and 5-quark SHBs where the inverse mass hierarchy holds in the 3-quark dominant SHBs, by adding $g_1'$ and $\mu_1'$ to the analysis in Sec.~\ref{sec:MixNoAnomaly}.

Now we have nine parameters of $\mu_1$, $\mu_1'$, $\mu_2$, $\mu_3$, $\mu_4$, $g_1$, $g_1'$, $h=A(g_2+g_3) + g_2'$ and $g_4$. First we use the masses of the four positive-parity SHBs as inputs to reduce the parameters. Next, we employ theoretically predicted $\Lambda_c(2890^*)$ and $\Xi_c(2765^*)$ as well as the experimentally observed $\Xi_c(2930)$ as other inputs. The input masses are summarized in Table.~\ref{tab:Parameter2}. In this table, $\Lambda_c(2890^*)$ which is 3-quark dominant is assigned to $\Lambda_c^H(-)$: $M[\Lambda_c^H(-)]=2890^*$ MeV, from the discussion around Eq.~(\ref{LambdaCOut}). Besides, for $\Xi_c(2765^*)$ and $\Xi_c(2930)$ obviously $M[\Xi_c^L(-)]=2765^*$ MeV and $M[\Xi_c^H(-)]=2930$ MeV from their mass ordering. Here, $\Xi_c(2765^*)$ is 3-quark dominant since the prediction is based on a three-quark picture, while $\Xi_c(2930)$ is 5-quark dominant to explain its small decay width as already explained. These properties forces us to impose another constraint that $m_{-,i=1,2}^{[{\bm 4}]}> m_{-,i=1,2}^{[{\bm 2}]}$, i.e., when the mixing disappears the mass of the 5-quark SHBs must be larger than that of the 3-quark SHBs. We note that the 3-quark dominant SHBs satisfy the inverse mass hierarchy $M[\Lambda_c(2890^*)]>M[\Xi_c(2765^*)]$, which is realized only when the $U(1)_A$ axial anomaly is present. From the inputs in Table.~\ref{tab:Parameter2}, seven parameters are fixed and only two parameters are left.

As for the remaining parameters, we take $\mu_1'$ and $g_4$ which are both responsible for the mixing strength. Depicted in Fig.~\ref{fig:LambdaMassAnomaly} demonstrates how the undetermined mass of 5-quark dominant $\Lambda_c^L(-)$ is constrained within our approach. In this figure the purple and green lines represent boundaries constrained from the decay widths of $\Xi_c(2970)$ and $\Xi_c(2930)$, respectively. That is, only the colored area enclosed by both the lines are allowed for the mass of $\Lambda_c^L(-)$. The resultant allowed mass is typically $M[\Lambda_c^L(-)]\sim 2700$ MeV, where the wide value of $\mu_1'$ stemming from the $U(1)_A$ anomaly effects is allowed. Therefore, even when the anomaly effects play a significant role so as to lead to the inverse mass hierarchy of the negative-parity 3-quark dominant SHBs, the mass of predicted 5-quark dominant $\Lambda_c(-)$ again lies approximately in the range of Eq.~(\ref{LambdaCOut}) whose decay width is of order a few MeV as estimated in Sec.~\ref{sec:DecayLambdaC}. It should be noted that points with $\mu_1'=0$ do not correspond to the absence of anomaly effects since $g_1'$ is always nonzero in the colored region.

\section{Discussion}
\label{sec:DecayLambdaC}

Our analysis in Sec.~\ref{sec:35Mixing} predicts the existence of a negative-parity 5-quark dominant $\Lambda_c(-)$ baryon whose mass is of order $2700$ MeV. However, within our present model where exact $SU(2)_h$ HQSS works, the predicted $\Lambda_c(-)$ baryon does not decay by the strong interaction. In this section, we incorporate a violation of $SU(2)_h$ HQSS to estimate the decay width of the $\Lambda_c(-)$.

The main decay mode of the 5-quark dominant $\Lambda_c(-)$ is expected to be $\Lambda_c(-)\to \Sigma_c\pi$, and here we evaluate its decay width. We note that this process is triggered by the violation of $SU(2)_h$ HQSS, since the spin and parity of the initial- and final-state diquarks are $0^-$ and $1^+$, respectively, and the one-pion-emission decay cannot preserve the light-spin conservation. The diquark $\tilde{d}^\mu$ as a building block of the HQS-doublet SHBs is Lorentz vector, and in the chiral basis $\tilde{d}^\mu$ takes the form of $(\tilde{d}_{ia}^\alpha)^\mu \sim \epsilon^{\alpha\beta\gamma}(q^T_L)^\beta_iC\gamma^\mu (q_R)^\gamma_a$ from the Pauli principle~\cite{Harada:2019udr}. That is, $\tilde{d}^\mu$ belongs to the $({\bm 3},{\bm 3})$ representation of $SU(3)_L\times SU(3)_R$ chiral symmetry, and accordingly the HQS-doublet SHBs $S^\mu_{ia}\sim Q^\alpha (\tilde{d}_{ia}^\alpha)^\mu$ transform as $S^\mu\to g_LS^\mu g_R^T$ under the chiral transformation. Hence, an interaction Lagrangian, ${\cal L}_{\rm HQSB}$, describing couplings among the HQS-doublet $S^\mu$, HQS-singlet $B'_{R(L)}$ and light mesons $\Sigma$, is obtained as 
\begin{eqnarray}
{\cal L}_{\rm HQSB}&=& \frac{\kappa}{2M_{\Lambda_c}}\epsilon_{\mu\nu\rho\sigma}\Big(\epsilon_{ijk}\bar{S}^{\mu}_{ai}\Sigma^{\dagger}_{aj}\Sigma^{\dagger}_{bk} v^{\nu}\sigma^{\rho\sigma}B^{\prime}_{L,b} \nonumber \\
&-& \epsilon_{abc}\bar{S}^{T\mu}_{ia}\Sigma_{ib}\Sigma_{jc} v^{\nu}\sigma^{\rho\sigma}B^{\prime}_{R,j}\Big) + {\rm h.c.}\ , \label{IntHQSViolation}
\end{eqnarray}
where $SU(3)_L\times SU(3)_R$ chiral symmetry is respected. In Eq.~(\ref{IntHQSViolation}), $\sigma^{\rho\sigma} = \frac{i}{2}[\gamma^\rho,\gamma^\sigma]$ is the antisymmetric Dirac matrix representing magnetic interactions and the minus sign of the second term stems from the parity invariance. It should be noted that we have defined the dimensionless coupling constant $\kappa$ by employing the mass of the ground-state $\Lambda_c$, $M_{\Lambda_c}=2286$ MeV, as a normalization factor in order to emphasize that the Lagrangian~(\ref{IntHQSViolation}) is HQSS-violating contributions. Besides, in Eq.~(\ref{IntHQSViolation}) we have introduced couplings involving only the 5-quark SHBs for the HQS-singlet, based on the fact that the 3-quark components in the $\Lambda_c(-)$ is small.

\begin{figure}[t]
\centering
\hspace{-0.5cm}
\includegraphics*[scale=0.46]{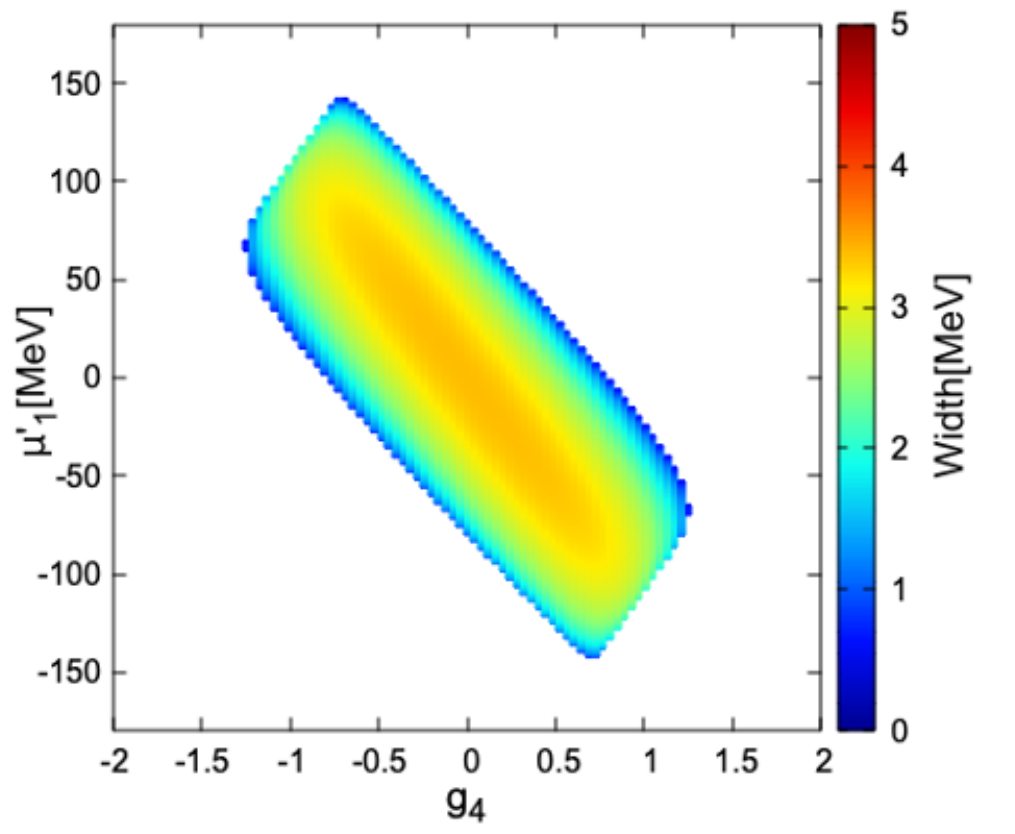}
  \caption{Decay width of $\Lambda_c(-)\to \Sigma_c\pi$ in the allowed region of $\mu_1'$ - $g_4$ plane.}
\label{fig:DecayHQSD}
\end{figure}

The $\Sigma_c$ baryons belong to ${\bm 6}$ representation of $SU(3)_{L+R}$ flavor symmetry carrying positive parities. More concretely $\Sigma_c$ baryons are described by replacing $S^\mu\to S^{6\mu}$ with the flavor-sextet SHB fields
\begin{eqnarray}
 S^{6\mu} =
\begin{pmatrix}
\Sigma^{I_3=1\mu}_c &\frac{1}{\sqrt2}\Sigma^{I_3=0\mu}_c &\frac{1}{\sqrt{2}}\Xi_c^{\prime I_3=\frac{1}{2}\mu} \\
    \frac{1}{\sqrt2}\Sigma^{I_3=0\mu}_c &\Sigma^{I_3=-1\mu}_c  &\frac{1}{\sqrt{2}}\Xi_c^{\prime I_3=-\frac{1}{2}\mu} \\
    \frac{1}{\sqrt{2}}\Xi_c^{\prime I_3=\frac{1}{2}\mu} &\frac{1}{\sqrt{2}}\Xi_c^{\prime I_3=-\frac{1}{2}\mu} &\Omega_c^{\mu} 
\end{pmatrix}. \label{S6Mu}
\end{eqnarray}
The spin $3/2$ and $1/2$ components of the HQS-doublet~(\ref{S6Mu}), $S^{6*\mu}$ and $S^6$, are separated by the following decomposition:
\begin{eqnarray}
S^{6\mu}_{ij}&=&S^{6\ast\mu}_{ij}-\frac{1}{\sqrt{3}}(\gamma^{\mu}+v^{\mu})\gamma_5S^{6}_{ij}\ . \label{SDecomposition}
\end{eqnarray}
Inserting Eqs.~(\ref{S6Mu}) and~(\ref{SDecomposition}) into the Lagrangian~(\ref{IntHQSViolation}) together with Eqs.~(\ref{NGBoson}) and~(\ref{Parity5q}), we can evaluate a decay width of $\Lambda_c(-)\to \Sigma_c\pi$. It should be noted that $\Lambda_c(-)\to\Sigma_c^*\pi$ is forbidden by the conservation of total angular momentum.

The magnitude of $\kappa$ in Eq.~(\ref{IntHQSViolation}) would be of ${\cal O}(1)$ as naturally expected. When we assume $\kappa=1$, the decay width of $\Lambda_c(-)\to \Sigma_c\pi$ can be estimated to be $1$ - $3$ MeV as shown in Fig.~\ref{fig:DecayHQSD} with the same setup of the analysis in Sec.~\ref{sec:MixWithAnomaly}. This value is substantially small compared to the widths of Roper-like SHBs whose total width is typically of order $50$ MeV. Such a small width reflects the fact that the decay processes violate $SU(2)_h$ HQSS.


\begin{figure*}[t]
\centering
\includegraphics*[scale=0.9]{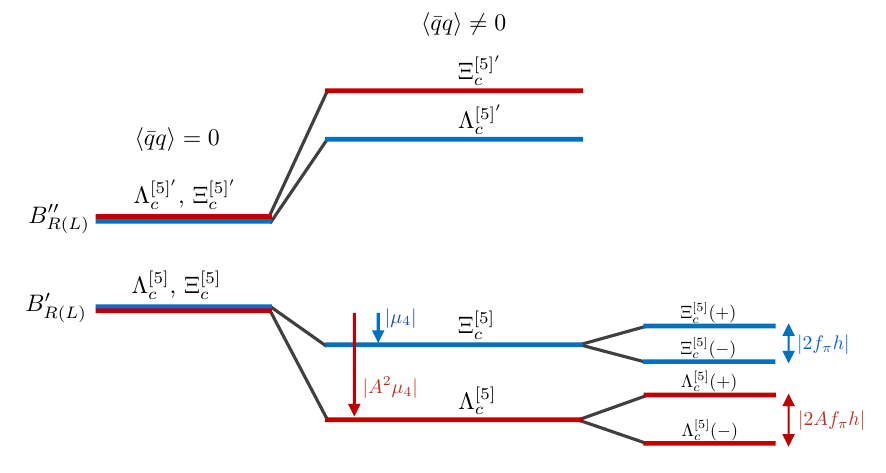}
  \caption{Schematic picture of generating the mass ordering of $M[\Lambda_c^{[5]}(\pm)] < M[\Xi_c^{[5]}(\pm)]$.}
\label{fig:Scenario}
\end{figure*}

\section{Conclusions}
\label{sec:Conclusions}

In this paper, we have investigated effects of the $U(1)_A$ axial anomaly on the mass spectrum of singly heavy baryons composed of three quarks ($Qqq$) and five quarks ($Qqq\bar{q}q$), based on the linear representation of $SU(3)_L\times SU(3)_R$ chiral symmetry. For pure 3-quark SHBs, we have shown that, the inverse mass hierarchy for negative-parity SHBs{\blue,} where the mass of $\Lambda_c$ becomes larger than that of $\Xi_c$ despite of quark contents, is triggered only when the $U(1)_A$ anomaly effects are present. In contrast, we have found that the anomaly effects do not have influence on a mass spectrum of SHBs containing pure 5-quark states at the leading order, and accordingly their decay properties are not affected.

When mixings between 3-quark and 5-quark SHBs are switched on, transitions between these two states become possible by emitting a pseudoscalar meson. Having focused on this feature, we have shown that the experimentally observed $\Xi_c(2923)$ or $\Xi_c(2930)$ can be a 5-quark dominant SHB, and its comparably small decay width is understood by a small mixing of the 3-quark SHB. As one of consequences of our present description, we have predicted the existence of a negative-parity 5-quark dominant $\Lambda_c$ baryon, mass and decay width of which are of order $2700$ MeV and a few MeV, respectively, regardless of the strength of the anomaly effects. Therefore, the predicted $\Lambda_c$ baryon is expected to provide a good experimental test of our picture for SHBs based on the conventional diquark ($qq$) and the tetra-diquark ($qq\bar{q}q$).

\section*{Acknowledgment}
The authors thank Kiyoshi Tanida for useful comments on experimental data of $\Xi_c$. D.S. is supported by the RIKEN special postdoctoral researcher program. This work is partially supported by the Japan Society for the Promotion of Science (JSPS) KAKENHI Grants, Nos.~23K03377 (D.S.), 20K03927, 23H05439 (M.H.), 18H05407, 21H04478 (A.H.), and 20K03959, 21H00132, 23K03427 (M.O.).

\appendix

\section{Possible interpretation of the unnatural ordering, $M[\Lambda_c^{[5]}(\pm)] < M[\Xi_c^{[5]}(\pm)]$}
\label{sec:Interpretation}

In Sec.~\ref{sec:5-quark}, we have found that the pure 5-quark SHBs satisfy the unnatural mass ordering $M[\Lambda_c^{[5]}(\pm)] < M[\Xi_c^{[5]}(\pm)]$ for both the positive-parity and negative-parity states although the quark contents of $\Lambda_c^{[5]}(\pm)$ and $\Xi_c^{[5]}(\pm)$ read $uds\bar{s}c$ and $uds\bar{u}c$ $(uds\bar{d}c)$, respectively. Such a peculiar mass ordering is mostly triggered by a negative contribution of $\mu_4$ term to their masses. In this appendix, we present a possible interpretation of obtaining $\mu_4<0$ by introducing couplings with excited 5-quark SHBs $B_{R(L)}^{\prime\prime}$. 

To begin with, we introduce the following orbitally excited tetra-diquarks as building blocks of $B_{R(L)}^{\prime\prime}$:
\begin{eqnarray}
(d^{\prime\prime}_{R})^\alpha_i&\sim&\epsilon_{jkl}\epsilon^{\alpha\beta\gamma}{(q^T_R)}^\beta_j C{(q_R)}^\gamma_k[{(\bar{q}_R)}^\delta_i\Slash{\partial}{(q_R)}^\delta_l]\ , \nonumber \\
(d^{\prime\prime}_{L})^\alpha_a&\sim&\epsilon_{bcd}\epsilon^{\alpha\beta\gamma}{(q^T_L)}^\beta_b C{(q_L)}^\gamma_c[{(\bar{q}_L)}^\delta_a\Slash{\partial}{(q_L)}^\delta_d]\ .
\end{eqnarray}
The chiral representation and $U(1)_A$ axial charge of $d_{R(L)}^{\prime\prime}$ read
\begin{eqnarray}
d^{\prime\prime}_{R}&\sim&(\bm 1,\bar{\bm 3})_{+2,}\ , \ \  d^{\prime\prime}_{L}\sim(\bar{\bm 3},\bm 1)_{-2}\ , 
\end{eqnarray}
which is distinct from those of $d_{R(L)}'$ in Eq.~(\ref{ChiralRep}) due to excitation properties stemming from $\Slash{\partial}$. The corresponding SHB fields are simply given by $B_{R(L)}^{\prime\prime} \sim Qd_{R(L)}^{\prime\prime}$. Thus, an interaction Lagrangian describing couplings among $B_{R(L)}'$, $B_{R(L)}^{\prime\prime}$ and $\Sigma$ which is invariant under $SU(3)_L\times SU(3)_R$ chiral transformation is obtained as
\begin{eqnarray}
{\cal L}_5^{\prime}
= - g_5(\bar{B}^{\prime}_{R}\Sigma^{\ast}B^{\prime\prime}_{R}+\bar{B}^{\prime\prime}_{L}\Sigma^{\ast}B^{\prime}_{L} + {\rm h.c.})\ . \label{5qIntEx}
\end{eqnarray}
In Eq.~(\ref{5qIntEx}) we have included only the leading order of $\Sigma^{(\dagger)}$ to see roles of the excited SHBs $B_{R(L)}^{\prime\prime}$ in a clear way. From Eq.~(\ref{5qIntEx}) classical equation of motions (EOMs) for $B_{R(L)}^{\prime\prime}$ are evaluated to be
\begin{eqnarray}
(i\Slash{\partial}-M_{5q}^{\prime\prime})B^{\prime\prime}_L=g_5\Sigma^{\ast}B^{\prime}_{L}\ , \nonumber\\
(i\Slash{\partial}-M_{5q}^{\prime\prime})B^{\prime\prime}_{R}=g_5\Sigma^{T}B^{\prime}_{R}\ , \label{EOM5qPP}
\end{eqnarray}
where $M_{5q}^{\prime\prime}$ denotes the mass of $B_{\pm}^{\prime\prime} \equiv (B_R^{\prime\prime}\mp B_L^{\prime\prime})/\sqrt{2}$ in the chiral symmetric phase. The kinetic terms in Eq.~(\ref{EOM5qPP}) can be neglected since the mass $M_{5q}^{\prime\prime}$ is much larger than the typical energy scale of QCD, $\Lambda_{\rm QCD}$. That is, the classical EOMs~(\ref{EOM5qPP}) are approximated to be
\begin{eqnarray}
B^{\prime\prime}_L=-\frac{g_5}{M_{5q}^{\prime\prime}}\Sigma^{\ast}B^{\prime}_{L}\ , \nonumber\\
B^{\prime\prime}_{R}=-\frac{g_5}{M_{5q}^{\prime\prime}}\Sigma^{T}B^{\prime}_{R}\ .\label{EOMRed}
\end{eqnarray}
Integrating out the heavier $B_{R(L)}^{\prime\prime}$ in Eq.~(\ref{5qIntEx}) with Eq.~(\ref{EOMRed}), we arrive at a reduced interaction Lagrangian of the form
\begin{eqnarray}
{\cal L}_5^{\prime} \sim \frac{g_5^2}{M_{5q}^{\prime\prime}}\Big[\bar{B}^{\prime}_{R}(\Sigma\Sigma^\dagger)^TB^{\prime}_{R}+\bar{B}^{\prime}_{L}(\Sigma^{\dagger}\Sigma)^TB^{\prime}_{L} \Big] \ . 
\end{eqnarray}
Therefore, comparing this expression with the $\mu_4$ term in Eq.~(\ref{L5q}), we can find
\begin{eqnarray}
\mu_4 \sim -\frac{g_5^2f_\pi^2}{M_{5q}^{\prime\prime}}<0\ ,
\end{eqnarray}
and the mass ordering of $M[\Lambda_c^{[5]}(\pm)] < M[\Xi_c^{[5]}(\pm)]$ is derived.

In the above derivation we have integrated out the excited $B_{R(L)}^{\prime\prime}$ to yield $M[\Lambda_c^{[5]}(\pm)] < M[\Xi_c^{[5]}(\pm)]$. Intuitively speaking, the mass ordering is driven by the level repulsion between $B_{R(L)}'$ and $B_{R(L)}''$ with the magnitude of $|\mu_4|$ for $\Xi_c^{[{\bm 5}]}$ and $|A^2\mu_4|$ for $\Lambda_c^{[{\bm 5}]}$, as depicted in Fig.~\ref{fig:Scenario}. In this figure, $\Lambda_c^{[{\bm 5}]'}$ and $\Xi_c^{[{\bm 5}]'}$ are the SHBs consisting of $d_{R(L)}''$. It should be noted that the $h$ contributions induced by ${\cal O}(\Sigma^4)$ terms and the $U(1)_A$ anomaly effects are expected to be of higher order as explicitly shown in Eq.~(\ref{PSet1}). Thus, the fine splitting of $\Xi_c^{[{\bm 5}]}(+)$ and $\Xi_c^{[{\bm 5}]}(-)$ or of $\Lambda_c^{[{\bm 5}]}(+)$ and $\Lambda_c^{[{\bm 5}]}(-)$ is relatively small as indicated in Fig.~\ref{fig:Scenario}.

\bibliography{reference}

\end{document}